\documentclass[a4paper,12pt]{article}
\title{LERW as an example of off-critical SLEs}
\date{}
\author{}

\newcommand{\vev}[1]{\langle #1 \rangle}
\newcommand{\bvev}[1]{\big\langle #1 \big\rangle}
\newcommand{\no}[1] {\, \mathrm{:} #1 \mathrm{:} \,}
\newcommand{\const} {\mathrm{const.} \,}
\newcommand{\im} {\Im \textrm{m }} % imaginary part
\newcommand{\bdry} {\partial}      % boundary
\newcommand{\eps} {\varepsilon}
\newcommand{\Order} {\mathcal{O}}
\newcommand{\bH} {\mathbb{H}}      % upper half plane
\newcommand{\bC} {\mathbb{C}}      % complex plane
\newcommand{\bR} {\mathbb{R}}      % real line
\newcommand{\bD} {\mathbb{D}}      % unit disc
\newcommand{\PR} {\mathsf{P}}      % {\mathsf{P}}
\newcommand{\EX} {\mathsf{E}}      % {\mathsf{E}}
\newcommand{\sF} {\mathcal{F}}     % filtrations..
\newcommand{\sL} {\mathcal{L}}     % loop erasure
\newcommand{\sZ} {\mathcal{Z}}     % ratio of partition functions
\newcommand{\sO} {\mathcal{O}}     % ``operator''
\newcommand{\lapl} {\triangle}     % Laplacian
\newcommand{\half} {\frac{1}{2}}
\newcommand{\minim} {\land}

\newcommand{\ind} {\mathbf{1}}     % indicator
\newcommand{\unit} {\mathbf{1}}    % unit operator
\newcommand{\bcond} {\mathrm{bdry} \; \mathrm{cond.}}
\newcommand{\cc} {\overline}       % complex conjugate
\newcommand{\ud} {\mathrm{d}}      % d for differential
\newcommand{\qv}[1]
    {\langle #1 \rangle}           % quadratic variation
\newcommand{\mint} {\int \! \! \! \cdot \! \! \cdot \! \! \cdot \! \! \! \int}

\newcommand{\lt} {\no{\chi^- \chi^+}}
\newcommand{\ltH} {\, \mathrm{:} \chi^- \chi^+ \mathrm{:}_\bH \,}
\newcommand{\ltD} {\, \mathrm{:} \chi^- \chi^+ \mathrm{:}_\bD \,}
\newcommand{\corrl} {\zeta}

\newcommand{\RW} {\mathrm{RW}}
\newcommand{\BM} {\mathrm{BM}}

\newcommand{\Bplan} {\bf{B}}

\def\debut{\begin{eqnarray}}
\def\fin{\end{eqnarray}}
\def\non{\nonumber}

\usepackage[final]{graphicx}
\usepackage{amssymb} % pour mathbb et mathfrak
\usepackage{amsmath}
\usepackage{oldgerm} % pour textgoth, textfrac et textswab
\usepackage{verbatim}
\input{epsf}

\begin{document}
\maketitle

\vspace{-1.5 truecm}

\centerline{\large Michel Bauer\footnote[1]{
    Service de Physique Th\'eorique de Saclay,
    CEA-Saclay, 91191 Gif-sur-Yvette, France and Laboratoire de 		Physique Th\'eorique, Ecole Normale Sup\'erieure, 24 rue 			Lhomond, 75005 Paris, France.
    {\small \tt <michel.bauer@cea.fr>}}, 
    Denis Bernard\footnote[2]{Member of the CNRS;
    Laboratoire de Physique Th\'eorique,
    Ecole Normale Sup\'erieure, 24 rue Lhomond, 75005 Paris, France.
    {\small \tt <denis.bernard@ens.fr>}},
    Kalle Kyt\"ol\"a\footnote[3]{
    Laboratoire de Physique Th\'eorique et Mod\`eles Statistiques,
    Universit\'e Paris Sud, 91405 Orsay, France and Service de 			Physique Th\'eorique de Saclay, CEA-Saclay, 91191 Gif-sur-Yvette, 	France. {\small \tt <kalle.kytola@u-psud.fr>}}}

\vspace{.3cm}

%\centerline{\large Service de Physique Th\'eorique de Saclay}
%\centerline{CEA/DSM/SPhT, Unit\'e de recherche associ\'ee au CNRS}
%\centerline{CEA-Saclay, 91191 Gif-sur-Yvette, France}

%\vspace{.3cm}

\vspace{1.0 cm}

\begin{abstract}
Two dimensional loop erased random walk (LERW) is a random curve,
whose continuum limit is known to be a Schramm-Loewner evolution (SLE)
with parameter $\kappa=2$.
In this article we study
``off-critical loop erased random walks'', loop erasures
of random walks penalized by their number of steps.
On one hand we are able to identify counterparts
for some LERW observables in terms of symplectic fermions ($c=-2$),
thus making further steps towards a field
theoretic description of LERWs. On the other hand, we show that
it is possible to understand the Loewner driving function of the
continuum limit of off-critical LERWs, thus providing an example of
application of SLE-like techniques to models near their critical point.
Such a description is bound to be quite complicated because outside the
critical point one has a finite correlation length and therefore no
conformal invariance. However, the example here shows the question need
not be intractable. We will present
the results with emphasis on general features that can be expected to
be true in other off-critical models.
\end{abstract}

% subject classification: Statistical mechanics.

%\vskip 1.5 truecm

\newpage

%\tableofcontents

\section{Introduction}

Over the last few years, our understanding of interfaces in two
dimensional systems at criticality has improved tremendously.
Schramm's idea \cite{Schramm-LERW_and_UST} to describe these
interfaces via growth processes has met a great success. We are now in
position to answer quantitatively in a routine way many questions of
interest for physicists and/or mathematicians (the overlap is only
partial, but non void).
 
All these successes suggest that we might try to be more ambitious and
it seems that time has come to start thinking about what can be said
for interfaces in non-critical systems.  So far, the only attempts in
this direction seem to be \cite{CFN-offcritical_percolation,
  NW-offcritical_percolation}, although also some yet unpublished work
\cite{MS-offcritical} will treat questions similar to this article in
various models.

Needless to say, we do not aim to achieve general and definitive
success in these notes. It is more our purpose to review some examples
and see what we can say in each situation. We shall go a bit deeper
into the specific example of loop erased random walks (LERW). This
choice has a number of reasons. First, certain quantities for the LERW
can be computed using only the underlying walk (with its loops kept),
whose scaling limit is the familiar Brownian motion. This is the case
for instance of boundary hitting probabilities. Second, the quantum
field theory of the LERW is that of symplectic fermions, a free
fermionic theory.  This relationship is part of the standard lore at
criticality, but it persists in the massive situation, and the
specific perturbation we study is related to the Brownian local time,
making it possible to compare closely the points of view of physics
and mathematics.
 
To understand the difficulties inherent to the study of noncritical
interfaces, it is perhaps worth spending some time on the physical and
mathematical views concerning criticality and conformal invariance

In statistical mechanics on the lattice, for generic values of the
parameters (collectively called $J$ here, examples include
temperature, pressure, magnetic field, fugacity) the connected
correlations among local observables decrease quickly (typically
exponentially) with the distance : a correlation length $n(J)$ (in
lattice units) can be defined and turns out to be of the order of a
finite number of lattice mesh. Achieving a large $n(J)$ requires to
adjust the parameters. Imagine we cover the plane (or approximate a
fixed domain in the plane) with a lattice of mesh $a$, and tune the
parameters $J$ in such a way that the macroscopic correlation length
$an(J)=\corrl$ remains fixed while $a$ goes to $0$. Then it is
expected on physical grounds that a limiting continuum theory exists,
which may describe only some of the initial degrees of freedom in the
system. Lattice translation symmetry becomes usual translation
invariance in the limit. Rotation invariance is also very often
restored. Over scales $s \gg  a$ the discrete system is expected to be
well approximated by the continuum theory.

In the limit $a \rightarrow 0$, $J$ tends to a limiting critical value
$J_c$ (when several parameters are present this may be a critical
manifold).  The approach of $J$ to $J_c$ when $a \rightarrow 0$ is
described by critical exponents. As $n(J_c)$ is infinite, so is
$an(J_c)$ for any $a$, and the macroscopic correlation length is
infinite at the critical point as well : the system has no
characteristic length scale %at the critical point
and the continuum limit is scale invariant. Over scales $s \ll \corrl$
the continuum off-critical system is expected to be well approximated
by the critical system. On the lattice, an infinite number of control
parameters can easily be exhibited, but if only their influence on the
long distance physics is considered the equivalence classes form
usually a finite dimensional space. Similarly, in the continuum limit,
usually only a finite number of perturbations out of criticality are
relevant.

For many two dimensional systems of interest, translation, rotation and
scale invariance give local conformal invariance for free : the 
descriptions of the system in two conformally equivalent geometries
are related by pure kinematics. This remarkable feature that emerges
only in the continuum limit is suggested by convincing physical
arguments but unproved in almost all cases of interest. The consequences of
conformal invariance have been vigorously exploited by physicists
for local observables since the 1984 and the seminal paper \cite{BPZ},
even if the road to a complete classification of local two dimensional
conformal field theories is still a distant horizon. 

Schramm's result in 1999 \cite{Schramm-LERW_and_UST}, on the other hand,
is a complete classification of probability measures on random curves
in (simply connected) domains of the complex plane (say joining two
boundary points for definiteness) satisfying two axioms : conformal
invariance and the domain Markov property. Again, the actual proof
that a lattice interface has a limiting continuum description which
satisfies the two axioms requires independent hard work.
However the number of treated cases is growing rapidly, including the
LERW, the Ising model, the harmonic navigator,
percolation\footnote{But it should be noted that most of the proofs
deal with a specific version on a specific lattice, which is rather
unsatisfactory for a physicist thinking more in termes of universality
classes.}.
A notorious exception which up to now has resisted to all attacks is
the case of self avoiding walks. 

Suppose that for each triple $({\mathbb D},x_0,x_\infty)$ consisting
of a domain with two marked boundary points one has a probability measure
on curves joining $x_0$
to $x_\infty$ (so we have the chordal case in mind). Consider an initial
segment of curve, say $\gamma$, joining $x_0$ to a bulk point $x'$ in
${\mathbb D}$. The domain Markov property relates the distribution
of random curves in two situations : it states the equality of 1)
the distribution of the rest of the random curve from $x'$ to $x_\infty$
in ${\mathbb D}$ conditional on $\gamma$ and 2) the distribution
of the random curve from $x'$ to $x_\infty$ in
${\mathbb D}\backslash \gamma$. 

This leads naturally to a description of the random curve as a growth
process : if one knows how to grow an (infinitesimal) initial segment
$\gamma$ in $({\mathbb D},x_0,x_\infty)$ from $x_0$ to $x_\infty$,
one can apply the
domain Markov property to build the rest of the curve as a curve
in the cut domain $({\mathbb D}\backslash \gamma,x',x_\infty)$
and then conformal invariance to "unzip" the cut i.e. map
$({\mathbb D}\backslash \gamma,x',x_\infty)$ conformally to
$({\mathbb D},x_0,x_\infty)$, so that another (infinitesimal) initial
segment can be grown and mapped back to
$({\mathbb D}\backslash \gamma,x',x_\infty)$
to get a larger piece of curve, and so on.  

Technically, Schramm's proof is made simpler by using the upper half
plane $\mathbb H$ with $0$ and $\infty$ as marked points, with a time
parameterization of the curve by (half) its capacity. Then the
conformal map $g_t(z)$ that unzips the curve grown up to time
$t$ and behaves at $\infty$ like $z+O(z^{-1})$ satisfies a
Loewner differential equation $\frac{dg_t(z)}{dt}=\frac{2}{g_t(z)-\xi_t}$
which amounts to encoding the growing curve via the real continuous
driving function $\xi_t$. It should be stressed that this representation
is valid for any curve (or more generally any locally growing hull),
independently of conformal invariance. However, the domain Markov
property and conformal unzipping of the random curves straightforwardly
translate into nice properties of the process $\xi_t$ : it has
independent and stationary increments. Continuity yields that
$\xi_t$ is a linear combination of a Brownian motion and time.
Finally scale invariance, the conformal transformations fixing
$({\mathbb H},0,\infty)$, leaves the sole possibility that
$\xi_t=\sqrt{\kappa}B_t$ for some normalized Brownian motion
$B_t$ and nonnegative scale factor $\kappa$. 

Now suppose we consider the system out of criticality. Intuitively,
there is no doubt that the probability that the interface has a
certain topology with respect to a finite number of points in the
domain should depend smoothly on the correlation length $\corrl$.
But can we say a bit more ? Conformal invariance cannot be used
to relate different domains and concentrating on the upper half
plane case, as in the following, is really a choice\footnote{Unless, 
as we will sometimes choose to do, we complicate matters 
by allowing the perturbation parameter (and thus correlation length) 
to vary from one point to another.}.
We can then describe the interface
again by a Loewner equation for a $g^\corrl_t(z)$ with some
(off-critical) source $\xi_t^\corrl$. What do we expect for this
new random process? 

At scales much smaller that the correlation length, i.e. in the
ultraviolet regime, the deviation from criticality is small, and
for instance the interface should look locally just like the
critical interface. This means that over short time periods,
the off-critical $\xi_t^\corrl$ should not be much different
from its critical counterpart. Is is easily seen that if
$\lambda >0$, the rescaled Loewner map
$\frac{1}{\lambda}g_{\lambda^2t}(\lambda z)$ still satisfies the
Loewner equation, but with a source
$\frac{1}{\lambda}\xi_{\lambda^2 t}^\corrl$. Taking a small
$\lambda $ amounts to zoom at small scales near the origin
and we expect that (in some yet unspecified topology)
$\lim_{\lambda \rightarrow 0^+}\frac{1}{\lambda}\xi_{\lambda^2 t}^\corrl$
exists and is a $\sqrt{\kappa}B_t$ for some normalized Brownian motion.
As the interface looks like a critical interface not only close to the
origin but close to any of its points, we also expect that
$g_s^\corrl$ maps the interface to a curve that looks like a critical
curve close to the origin, so that more generally for fixed $s$ the
limit
$\lim_{\lambda \rightarrow 0^+}\frac{1}
{\lambda}(\xi_{s+\lambda^2 t}^\corrl-\xi_{s}^\corrl)$
should exist and be a $\sqrt{\kappa}B_t$. Hence to each fixed $s$
we can in principle define a Brownian motion. The Brownian motions
defined for distinct $s$'s are moreover expected to be independent.
To go further, we would need to have some control on how uniform in
$s$ the convergence is, and how fast the correlations between
$\frac{1}{\lambda}(\xi_{s+\lambda^2 t}^\corrl-\xi_{s}^\corrl)$
for distinct values of $s$ decrease with $\lambda$. There could
be some problem with inversions of limits. In the nice situation,
we would naively deduce for the above facts that the quadratic
variation of $\xi_t^\corrl$ is exactly $\kappa t$ even at finite
$\corrl$.
This raises the question whether $\xi_t^\corrl$ can be represented
as the sum of a Brownian motion (scaled by $\sqrt{\kappa}$) plus
some process, contributing $0$ to the quadratic variation, but
whose precise regularity would remain to be understood. Finally,
the strongest relationship one could imagine between $\xi_t^\corrl$
and its critical counterpart $\xi_t$ would be that their laws are
mutually absolutely continuous over finite time intervals\footnote{As 
a consequence, we may expect a Radon-Nikodym derivative for the 
interface at best in finite domain but not in infinite domain such as
the half plane.}. We shall see examples of this situation in the sequel, 
but at least one counterexample is known, off-critical percolation
\cite{NW-offcritical_percolation}. On the lattice, the set of
interfaces is discrete, and the question of absolute continuity
trivializes. One can write down discrete martingales describing
the relative weight of an initial interface segment off/at
criticality and a naive extrapolation to the continuum limit yields
a candidate for the Radon-Nikodym derivative for the growth of
the interface. This is the basis of much of the forthcoming
discussion.  

At scales large with respect to $\corrl$ however, i.e. in the
infrared regime, the behavior is different and the interface
should look like 
another SLE with a new $\kappa _{ir}$. Think of the Ising model
for example. At criticality $\kappa=3$ but if the temperature
is raised above the critical point, renormalization group
arguments indicate that at large scale the interface looks
like the interface at infinite temperature, i.e; percolation
and $\kappa _{ir}=6$. One expects in general that
$\lim_{\lambda \rightarrow +\infty}\frac{1}
{\lambda}\xi_{\lambda^2 t}^\corrl$ exists and is a
$\sqrt{\kappa _{ir}}B_t$. This means that the process
$\xi_{t}^\corrl$ could yield information on the  flow of the
renormalization group. Whether this can be used as an
effective tool is unclear at the moment.

Let us close this introduction with the following observations.
Conformal invariance and the domain Markov property have a rather
different status. Whereas conformal invariance emerges (at best)
in the continuum limit at criticality, the domain Markov property
makes sense and is satisfied on the lattice without tuning
parameters for many systems of interest. It can be considered
as a manifestation of locality (in the physicists terminology).
Hence the domain Markov property is still expected to hold off
criticality. However the consequences of this property on
$\xi_t^\corrl$ do not seem to have a simple formulation. As
for conformal invariance, there is a trick to preserve it
formally out of the critical point : instead of perturbing
with a scaling field $O(z,\bar{z})$ times a coupling
\textit{constant} $\lambda$, perturb by a scaling field times a
density $\lambda(z,\bar{z})$ of appropriate weight, in such a
way that $\lambda (z,\bar{z}) O(z,\bar{z}) d\bar{z}\wedge dz$
is a $2$-form. This also gets rid of infrared divergences that
occur in unbounded domain if $\lambda(z,\bar{z})$ has compact
support. We shall use this trick in some places, but beware
that if perturbation theory contains divergences, problems
with scale invariance will arise, hence the cautious word
 "formally" used above.

\bigskip

The paper is organized as follows. We start with a few examples
in section \ref{sec: examples}. A brief account of SLEs, as
appropriate for our needs, is given in section \ref{sec: SLE basics}.
Section \ref{sec: probability measures} is devoted to the general
philosophy of how one might
hope to attack the question of interfaces in off-critical models.
In particular we propose a field theoretical formula for
Radon-Nikodym derivative between the off-critical and critical
measures on curves.
The main example of LERW is treated in detail in section
\ref{sec: LERW}. We discuss the critical and off-critical field
theory for LERW, compute multipoint functions of the perturbing
operator and subinterval hitting probabilities --- and derive
in two ways the off-critical driving process to first order in
the magnitude of the perturbation.

\section{Examples}
\label{sec: examples}

To give some concreteness to the thoughts presented in the introduction,
we will start with a couple of examples.

\subsection{Self avoiding walks}
Our first example deals with self avoiding walks (SAW). 
Consider a lattice of mesh size $a$ embedded in a domain 
$\mathbb{D}$ in the complex plane. A sample of a SAW is a 
simple nearest neighbor path on the lattice never visiting twice any
lattice site.
The statistics of SAW is specified by giving the weight 
$w_\gamma= x^{|\gamma|}$, with $|\gamma|$ the number of steps of 
$\gamma$ and $x$ the fugacity, to each path $\gamma$. 
The partition function $Z_\mathbb{D}$ is
the sum $Z_\mathbb{D}=\sum_\gamma x^{|\gamma|}$ and the probability of
occurrence of a curve $\gamma$ is $w_\gamma/Z_\mathbb{D}$.

There is a critical value $x_c$, depending on the lattice, for which 
typical sample consists of paths of macroscopic sizes so that the
continuum limit $a\to0$ can be taken. This continuum limit is conjectured 
to be conformally invariant and described by chordal SLE$_{8/3}$
if we restrict ourselves to SAW starting and ending at prescribed points
on the boundary of $\mathbb{D}$.

The off-critical SAW model in the scaling regime consists of looking
at SAW for fugacity $x$ close to its critical value $x_c$ (and approaching 
$x_c$  in a appropriate way as the mesh size goes to zero). The continuous
limiting theory is not anymore conformally invariant as a scale parameter is
introduced when specifying the way $x$ approaches its critical value.
Renormalization group arguments tell us that if $x<x_c$ the fugacity
flows to zero 
at large distances so that the partition function
is dominated by the shortest path
while if $x>x_c$ it flows to the critical value corresponding to
uniform spanning trees (UST) 
so that the partition function is dominated by these space filling paths.

The off-critical partition function
$\sum_\gamma x_c^{|\gamma|} (x/x_c)^{|\gamma|}$
can be written as an expectation value with respect to the critical measure
$$Z_\bD/Z^c_\bD= \EX[\, (x/x_c)^{|\gamma|} \,]$$
with $Z^c_\bD$ the critical partition function and $\EX$ the critical measure. 
If a sample path $\gamma$ has a typical length scale $l_\bD(\gamma)$, 
which is macroscopic in the critical theory, its number of steps scales as 
$|\gamma|\simeq (l_\bD(\gamma)/a)^{d_\kappa}$ 
with $a$ the lattice mesh size and $d_\kappa$ the fractal dimension
(more or less by definition of the fractal dimension). 
The scaling limit defining the continuous off-critical theory then consists of
taking the limit $x\to x_c$ such that $\nu:=-a^{-d_\kappa} \log(x/x_c) $
is finite 
as $a\to 0$, that is $(x-x_c)/x_c\simeq -\nu\,a^{d_\kappa}$ as $a\to0$. The 
parameter $\nu$ has scaling dimension $d_\kappa$ and introduces a scale
and a typical correlation length $\corrl\simeq \nu^{-1/d_\kappa}$.
This ensures that the relative weights 
$(x/x_c)^{|\gamma|}\simeq e^{-\nu |\gamma| a^{d_\kappa}} $ have 
a finite limit for typical paths as the mesh size goes to zero.
In the continuum the weights (relative the critical weights) 
are $e^{-\nu {L}_\bD(\gamma)}$
with ${L}_\bD(\gamma)\simeq l_\bD(\gamma)^{d_\kappa}$
and the ratio of the off-critical partition function to the critical
one is
$$\sZ_\bD= \EX[\, e^{-\nu {L}_\bD(\gamma)} \,].$$

In the continuum limit the critical curve should be
described by SLEs and we'd like to understand what the above teaches us
about the off-critical curves.
Recall that SLE comes
as a one parameter family SLE${}_\kappa$, $\kappa > 0$, and that
critical SAW is conjectured to be described by SLE$_{8/3}$.
For simplicity we consider here chordal SLE in which one looks 
for curves starting and ending at fixed points $x_0,\ x_\infty$
on the boundary $\bdry \bD$ of $\bD$. 
Recall that in the SLE construction the curves are given
a 'time' parametrization 
$\gamma:[0,T] \rightarrow \overline{\bD}$,
with $\gamma_0=x_0$, $\gamma_T=x_\infty$,
such that the filtration associated to the knowledge 
of the curve up to time $t$, $(\sF^\gamma_t)_{t\in[0,T]}$, is the
filtration generated by the Loewner driving process $(\xi_t)_{t\in[0,T]}$,
i.e. $\sF^\gamma_t = \sigma \{ \xi_s : 0 \leq s \leq t \}$
(for a reader not yet familiar with SLE, see section \ref{sec: SLE basics}).
The expectation $\EX[\cdots]$ in the previous formula becomes the SLE measure.
The mathematical definition of ${L}_\bD(\cdot)$ is related to what
is known as natural parametrization of the SLE curve
\cite{natural_parametrization1,natural_parametrization2, Kennedy-seminar}.
It should satisfy the additivity property
$${L}_\bD(\gamma_{[0,t+s]})=
{L}_\bD(\gamma_{[0,t]})+{L}_{\bD\setminus\gamma_{[0,t]}}(\gamma_{[t,t+s]}),$$
or even the stronger property that the natural parameterization of a
piece of SLE can be
defined without reference to the domain, and ${L}(\gamma_{[0,t+s]})=
{L}(\gamma_{[0,t]})+{L}(\gamma_{[t,t+s]}),$
since ${L}(\gamma)$ is naively proportional to the number of steps of
$\gamma$. We shall later define in a more general context the notion
of an interface energy and see that it
possesses an analogous additivity property. 

The factor $e^{-\nu {L}_\bD(\cdot)}$ specifies the Radon-Nikodym derivative
of the off-critical measure with respect to the critical SLE measure
so that the off-critical expectation of an observable ${\cal O}$ is 
$$ \EX^\nu[\, {\cal O} \,] = \sZ_\bD^{-1}\,
    \EX[\, e^{-\nu {L}_\bD(\cdot)}\, {\cal O}\,]
\quad {\rm with}\quad \sZ_\bD=\EX[\, e^{-\nu {L}_\bD(\cdot)}\,].$$
If ${\cal O}$ is $\sF^\gamma_t$-measurable, that is if ${\cal O}$
only depends on the
knowledge of the curve up to time $t$, we have
$$ \EX^\nu[\, {\cal O} \,] = \EX[\, M_t\, {\cal O}\,] \quad {\rm with} \quad
M_t:= \sZ_\bD^{-1}\, \EX[e^{-\nu {L}_\bD(\cdot)}|\sF^\gamma_t]$$
since $\EX[\, e^{-\nu {L}_\bD(\cdot)}\, {\cal O}]
=\EX[\, \EX[\,e^{-\nu {L}_\bD(\cdot)}\, {\cal O}|\sF^\gamma_t\,]\,]
= \EX[\, M_t\, {\cal O}\,] $ because
we can take out what is known,
$\EX[\, e^{-\nu {L}_\bD(\cdot)}\, {\cal O}|\sF^\gamma_t\,]=
{\cal O}\, \EX[\, e^{-\nu {L}_\bD(\cdot)}|\sF^\gamma_t\,]$.

In other words, the off-critical SLE (corresponding to the perturbation by
the natural parametrization) is obtained by weighting the SLE expectation with
$M_t$. As a conditional expected value,
$M_t$ is by construction a martingale and
$M_0=1$ so that $\EX^\nu$ is correctly normalized to be a probability
measure. Notice that, modulo a few regularity assumptions, this
is the framework 
in which Girsanov's theorem applies, as will be discussed in section
\ref{sec: Girsanov}. The additivity property of the natural 
parametrization implies that $M_t= e^{-\nu {L}_\bD(\gamma_{[0,t]})}\,
\sZ_\bD^{-1}\,\EX[e^{-\nu {L}_{\bD\setminus\gamma_{[0,t]}}(\cdot)}]$
so that
\begin{eqnarray}
 M_t= e^{-\nu {L}_\bD(\gamma_{[0,t]})}\, 
\frac{ \sZ_{\bD\setminus \gamma_{[0,t]}} }{\sZ_\bD}
\textrm{ .}
\label{Msaw}
\end{eqnarray}
$M_t$ can naturally be interpreted as the off-critical weight (relative
to the critical one) given to the curve $\gamma_{[0,t]}$. 
It is made of two contributions,
one is a ratio of partition functions in the cut domain 
$\bD\setminus \gamma_{[0,t]}$ and in $\bD$, the other is 
an 'interface energy' contribution $\nu {L}_\bD(\gamma_{[0,t]})$ 
associated to the curve.
We shall recover this decomposition in a more general (but more formal) 
context of perturbed SLEs in following sections.

\subsection{Loop erased random walks}
The second example deals with loop erased random walks (LERW) and
it will be further developed in section \ref{sec: LERW}.
Let us first recall the definition of a LERW. Again let us start with
a lattice $\bD^{(a)}$ of mesh $a$ embedded in a domain $\bD$. 
Given a path $W=(W_0,W_1,\cdots,W_n)$ on the lattice
its loop erasure $\gamma$ is defined as follows:
let $n_0=\max \{ m : W_m = W_0 \}$ and
set $\gamma_0=W_{n_0}=W_0$, next let 
$n_{1}=\max \{ m : W_m = W_{n_0+1} \}$ and set
$\gamma_1=W_{n_1}$, and then inductively let
$n_{j+1}=\max \{ m : W_m = W_{n_j+1} \}$ and set
$\gamma_j=W_{n_j}$. This produces a simple path
$\gamma = \sL(W) = (\gamma_0, \gamma_1, \ldots, \gamma_l)$
from $\gamma_0=W_0$ to $\gamma_l=W_n$,
called the loop-erasure of $W$, 
but its number of steps $l$ is in general much smaller 
than that of the original path $W$. 
We emphasize that the starting and end points are not changed
by the loop-erasing.

We point out that the above definition of loop erasure is equivalent to
the result of a recursive procedure of chronological loop erasing:
the loop erasure of a $0$ step path
$(W_0)$ is itself, $\gamma = (W_0)$ and if the erasure of
$W = (W_0,\ldots,W_m)$ is the simple path
$\sL(W) = (\gamma_0, \ldots, \gamma_l)$ then for the loop erasure
of $W' = (W_0, \ldots, W_m, W_{m+1})$ there are two
cases depending on whether a loop is formed on step $m+1$.
If $W_{m+1} \notin \{ \gamma_0 , \ldots, \gamma_l\}$ then the
loop erasure of $W'$ is
$\gamma' = (\gamma_0, \ldots, \gamma_l, W_{m+1})$. But if
a loop is formed, $W_{m+1} = \gamma_k$ for some $k \leq l$
(unique because $\gamma$ is simple), then the loop erasure
of $W'$ is $\gamma' = (\gamma_0, \ldots, \gamma_k)$.

In this paper we shall be interested in paths starting at a
boundary point ${x_0}$
and ending on a subset $S$ of the boundary of $\bD$.

Statistics of LERW is defined by associating to any simple path
$\gamma$ a weight 
$w_{\gamma} = \sum_{W : \sL(W) = \gamma} \mu^{|W|}$, where the sum is
over all nearest neighbor paths $W$ whose erasures produce $\gamma$, and
$|W|$ denotes the number of steps of $W$. 
There is a critical value $\mu_c$ of the fugacity at 
which the underlying paths $W$ become just ordinary random walks.
The partition function $\sum_\gamma w_\gamma$ of LERWs from
$z$ to $S$ in $\bD$ can be rewritten as a sum over walks
in the domain $\bD$, started from $z$ and counting only those
that exit the domain through set $S$
\debut \non
Z^{\bD ; z ; S}_{\RW}
\; = \; \sum_{\substack{\gamma \textrm{ simple path} \\
    \textrm{from $z$ to $S$ in $\bD$}}} w_\gamma
\; = \; \sum_{\substack{W \textrm{ walk from}\\Ê  
\textrm{$z$ to $S$ in $\bD$}}} \mu^{|W|}
\textrm{ .}
\fin
Written in terms of critical random walks, the 
partition function thus reads
$\EX^{z}_{\RW} \big[ (\mu/\mu_c)^{|W|} \;
    \ind_{W_{\tau^\RW_\bD} \in S} \big]$,
where
$\tau^{\RW}_\bD$ denotes the exit time of the random walk $W$ from $\bD$.

Critical LERW corresponds to the critical fugacity and is described
by SLE$_2$, see \cite{Schramm-LERW_and_UST, LSW-LERW_and_UST,
Zhan-scaling_limits_of_LERW}.
For $\mu<\mu_c$ --- which is the case we shall consider --- 
paths of small lengths are more favourable and
renormalization group arguments tell that at large distances the
path of smallest length dominates. 
The off-critical theory in the scaling regime corresponds to non
critical fugacity 
$\mu$ but approaching the critical one as the mesh size tends to zero.
At fixed typical macroscopic size, the number of steps of typical
critical random walks 
(not of their loop erasures) scales as $a^{-2}$, so that the
scaling limit is such
that $\nu:=-a^{-2} \log(\mu/\mu_c)$ is finite as $a\to0$, ie. 
$(\mu-\mu_c)/\mu_c\simeq -\nu\, a^2$ and $\nu$ has scaling dimension $2$
and fixes a mass scale $m\simeq \sqrt{\nu}$ and a correlation length 
$\corrl\simeq 1/m$. In this scaling limit the weights become 
$(\mu/\mu_c)^{|W|}\simeq e^{-\nu a^2|W|}$ and the random walks converge
to two dimensional Brownian motions $\Bplan$ with $a^2 |W| = a^2 \tau^\RW_\bD$
converging to
the times $\tau_\bD$ spent in $\bD$ by
%the two dimensional Brownian motions
$\Bplan$ before exiting.
The off-critical partition function 
can thus be written as a Brownian expectation value
$Z^{\bD; z; S}_{\nu} \longrightarrow
\EX^{z}_{\BM} \big[\, e^{-\nu \tau_\bD} \,
\ind_{\Bplan_{\tau_\bD} \in S} \, \big]$ as $a \downarrow 0$.
We may generalize this by letting $\nu$ vary in space:
steps out of site $w \in \bD$ are given weight factor
$\mu(w) = \mu_c \; e^{-a^2 \nu(w)}$ , in which case the partition
function is
\debut \non
Z^{\bD; z; S}_{\nu} =
    \EX^{z}_{\RW} \big[\,
    e^{-\sum_{0 \leq j < \tau^{\RW}_\bD} a^2 \nu(W_j)} \,
    \ind_{W_{\tau^{\RW}_\bD} \in S} \, \big]
\underset{a \downarrow 0}{\longrightarrow}
    \EX^{z}_{\BM} \big[\, e^{-\int_0^{\tau_\bD} \nu({\Bplan}_s) \, \ud s} \,
    \ind_{{\Bplan}_{\tau_\bD} \in S} \, \big]
\textrm{ .}
\fin
The explicit weighting by $e^{-\nu \tau^\RW_\bD}$ is transparent for the
random walk, but becomes less concrete for the LERW since the same
path $\gamma$ can be produced by random walks of different lengths and
by walks that visit different points.
Compared to the previous example of SAW, the description of the
off-critical LERW theory via SLE martingales
is thus more involved but will (partially) be 
described in following sections.

\subsection{Percolation}

We will briefly also mention the case of off-critical percolation,
just for some comparisons.
A way to study the scaling limit in the off-critical regime was
suggested in \cite{CFN-offcritical_percolation}.

It is most convenient to define interfaces in percolation on the
hexagonal lattice. A configuration $\omega$ of face percolation on
lattice domain $\bD$ with lattice spacing $a$ is
a colouring of all faces (hexagons) to open (1) or closed (0),
i.e. $\omega \in \{ 0,1 \}^{F_\bD}$, where $F_\bD$ is the set of
faces. The Boltzmann weight of a configuration is
$w_\omega = p^{\# \mathrm{open}(\omega)}
(1-p)^{\# \mathrm{closed} (\omega)}$ --- that is
hexagons are chosen open with probability $p$ independently.
This hexagonal lattice face percolation (triangular lattice site
percolation) is critical at $p=p_c=1/2$ and has been proven to be
conformally invariant \cite{Smirnov-percolation}. Exploration path,
an interface between closed and open clusters, is described in the
continuum limit by SLE${}_6$.
Note also that
$$Z_\bD = \sum_\omega p^{\# \mathrm{open}(\omega)}
    (1-p)^{\# \mathrm{closed}(\omega)}
= \prod_{z \in F_\bD} (p + (1-p)) =  1$$
for any $p \in (0,1)$.

The off-critical regime now consists of changing the state of
some faces that are macroscopically pivotal, i.e. affect
connectivity properties to macroscopic distances. Informally,
these faces are such that from their neighborhood there exists four
paths of alternating colors to a macroscopic
distance away from the point. The number of such points in the
domain $\bD$ should be of order $|\bD| \times a^{-3/4}$, so that
in order to have finite probability of changing a macroscopically
pivotal face we should take $|p-p_c| \sim a^{3/4}$.
We denote the perturbation amplitude by
$\nu = a^{-3/4} \log (\frac{p}{p_c})$.

For more about interfaces in off-critical percolation,
the reader should turn to
\cite{CFN-offcritical_percolation, NW-offcritical_percolation}.
The above remarks will be enough for us to give a point of
comparison.

\section{SLE basics}
\label{sec: SLE basics}

The method of Schramm-Loewner evolutions (SLE) is a significant
recent development in the understanding of conformally
invariant interfaces in two dimensions. We will describe the main
ideas briefly and informally, and refer the reader to the many
reviews of the topic, e.g.
\cite{Lawler-conformally_invariant_processes, Werner-lectures,
KN-guide_to_SLE, BB-2d_growth_processes, Cardy-SLE_for_physicists},
among which one can choose according to the desired level
of mathematical rigour, physical intuition, emphasis and
prerequisite knowledge.

\subsection{Chordal SLE in the standard normalization}

It was essentially shown in \cite{Schramm-LERW_and_UST} that
with assumptions of conformal invariance and domain Markov
property, probability measures on random curves in a simply connected domain $\bD$
from a point $x_0 \in \bdry \bD$ to $x_\infty \in \bdry \bD$ are
classified by one parameter, $\kappa \geq 0$. These random
curves are called chordal SLE${}_\kappa$.

The curves SLE${}_\kappa$ are simple curves (no double points)
iff $0\leq \kappa \leq 4$. For the purposes of this paper
simple curves are enough, so we restrict ourselves to this least
complicated case.
To describe the chordal SLE${}_\kappa$, we note that by the assumed
conformal invariance it suffices to discuss it 
in the domain $\bH = \{ z \in \bC : \im z > 0 \}$ (upper half plane)
from $0$ to $\infty$ --- for any other choice of $\bD, {x_0}, {x_\infty}$ one
applies a conformal map $f:\bH \rightarrow \bD$ such that
$f(0)=x_0$, $f(\infty)=x_\infty$. The existence of such $f$ follows from
Riemann mapping theorem and well-definedness of the resulting curve
($f$ is only unique up to composition with a scaling
$z \mapsto \lambda z$ of $\bH$)
from the scale invariance of chordal SLE.

So, let $0 \leq \kappa \leq 4$ and let $g_t(z)$ be the solution of
the Loewner's equation
\debut \label{eq: Loewner}
\frac{\ud}{\ud t} g_t(z) = \frac{2}{g_t(z)-\xi_t}
\fin
with initial condition $g_0(z) = z \in \bH$ and
$\xi_t = \sqrt{\kappa} B_t$ a Brownian motion with variance parameter
$\kappa$. The solution exists up
to time $t$ for $z \in \bH \setminus \gamma[0,t]$, where
$\gamma:[0,\infty] \rightarrow \overline{\bH}$
is a random simple curve such that
$\gamma_0 = 0$ and
$\gamma_\infty = \infty$. This curve is called
the (trace of) chordal SLE${}_\kappa$. Furthermore, $g_t$ is the
unique conformal map from $\bH \setminus \gamma[0,t]$ to $\bH$ with
the hydrodynamic normalization
$g_t(z) = z + \sO (z^{-1})$ as $z \rightarrow \infty$.

\subsection{Chordal and dipolar SLEs in the half plane}
The Loewner's equation (\ref{eq: Loewner}) can be used to describe
any simple curve 
in $\bH$ starting from the boundary $\bdry \bH = \bR$
in the sense that $g_t$ is the hydrodynamically normalized conformal
map from the complement of an initial segment of the curve to the half plane.
In particular, a chordal SLE${}_\kappa$ in $\bH$ from $x_0 \in \bR$ to
$x_\infty \in \bR$ is obtained by letting $\xi_0={x_0}$,
$\eta_0={x_\infty}$ and
$\xi_t$, $\eta_t$ solutions of the It\^o differential equations
\debut \non
\begin{cases}
\ud \xi_t \; = \; \sqrt{\kappa} \; \ud B_t
    + \frac{\rho_c}{\xi_t - \eta_t} \; \ud t & \\
\ud \eta_t \; = \; \frac{2}{\eta_t - \xi_t} \; \ud t
    & \qquad \textrm{ that is } \eta_t = g_t(x_\infty)
\end{cases}
\fin
with $\rho_c = \kappa-6$, see e.g.
\cite{SW-SLE_coordinate_changes, BBK-multiple_SLEs}.
The maximal time interval of the solution is
$t \in [0,T]$, where $T$ is a (random) stopping time and
$\gamma_T = x_\infty$.

Another interesting case is a curve in $\bH$ depending on the starting
point $x_0$ and two other points $x_+ < x_-$ such that $x_0 \notin [x_+,x_-]$.
If the two points play a symmetric role, then the appropriate random
conformally invariant curve is the dipolar SLE${}_\kappa$
\cite{BBH-dipolar_SLE}. We again
have the Loewner's equation (\ref{eq: Loewner}) with $\xi_0 = {x_0}$,
$X^\pm_0 = x_\pm$ and It\^o differential equations
\debut \non
\begin{cases}
\ud \xi_t \; = \; \sqrt{\kappa} \; \ud B_t
    + \frac{\rho_d}{\xi_t - X^+_t} \; \ud t
    + \frac{\rho_d}{\xi_t - X^-_t} \; \ud t & \\
\ud X^\pm_t \, = \, \frac{2}{X^\pm_t - \xi_t} \; \ud t
    & \qquad \textrm{ that is } X^\pm_t = g_t(x_\pm)
\end{cases}
\fin
with $\rho_d = \frac{\kappa-6}{2}$. Again dipolar SLE is defined for
$t \in [0,T]$, where $T>0$ is (random) stopping time such that
$\gamma_T \in [x_+,x_-]$.

Both these examples can be understood from the point of view of
statistical physics in such a way that the (regularized) partition
function for the model in $\bH$ is
$Z(x_0,x_\infty) = |x_\infty-x_0|^{\rho_c/\kappa}$
for the chordal SLE${}_\kappa$
and $Z(x_0,x_+,x_-) = |x_- - x_0|^{\rho_d/\kappa} \;
|x_+ - x_0|^{\rho_d/\kappa} \;
|x_- - x_+|^{\rho_d^2 \, / \, 2 \kappa}$ for the dipolar SLE${}_\kappa$.
The driving process satisfies
$\ud \xi_t = \sqrt{\kappa} \; \ud B_t + \kappa (\partial_\xi \log Z) \; \ud t$
and other points follow the flow $g_t$.
For discussion of more general SLE variants of this kind see
\cite{BBK-multiple_SLEs, Kytola-SLE_kappa_rho, Kytola-local_mgales,
BB-2d_growth_processes}.

\section{Probability measures}
\label{sec: probability measures}

\subsection{Definition from discrete stat. mech. models}
Let us first recall how measures on curves are defined in statistical
physics models  via Boltzmann weights. We have in mind Ising like models.
Let $\mathcal{C}$ be the configuration 
space of a lattice statistical model defined on a
domain $\mathbb{D}$. For simplicity we assume $\mathcal{C}$ to be
discrete and finite but as large as desired. 
Let $w_c$, $w_c \geq 0$, $c\in{\cal C}$, be the
Boltzmann weights and $Z_\mathbb{D}$ the partition function,
$Z_\mathbb{D}:=\sum_{c\in\mathcal{C}}\ w_c$.
By Boltzmann rules, the probability of a configuration $c$ is
$\PR[\{c\}]:= w_c/Z_\mathbb{D}$, and this makes $\mathcal{C}$
a probability space. 

In the present context, imagine that specific boundary conditions 
are imposed in such a way as to ensure the presence of an interface in $\bD$ 
for any sample  -- for simplicity we consider only one interface. 
Given a curve $\gamma$ in $\bD$, that we aim 
at identifying as an interface, there exists a subset of configurations
$\mathcal{C}_{\gamma}$ for which the actual interface
coincides with the prescribed curve $\gamma$.
Again by Boltzmann rules, the probability of occurrence of the
curve $\gamma$ as an interface, i.e. the probability of the event 
$\mathcal{C}_{\gamma}$, is the ratio of the partition functions
\begin{eqnarray}
\PR_\bD[\mathcal{C}_{\gamma}]= Z_\mathbb{D}[\gamma]/Z_\mathbb{D}.
\label{latticeproba}
\end{eqnarray}
where $Z_\mathbb{D}[\gamma]$ is the conditioned partition function 
defined by the restricted sum
$$
Z_\mathbb{D}[\gamma]:=\sum_{c\in\mathcal{C}_{\gamma}} w_c~.
$$

The Boltzmann weights may depend on parameters such that for critical values
the statistical model is critical. We denote by $\PR_\bD^0$ the
probability measure at criticality, 
with Boltzmann weight $w^0$,  which in the continuum is expected
to become an SLE measure if only the statistics of the interface
are considered. 
We generically denote by $\PR_\bD^\nu$ 
the off-critical measures, with Boltzmann weights $w^\nu$. These probability
measures differ by a density:
$$ \PR_\bD^\nu = M^\nu_\bD\  \PR_\bD^0$$
where, by construction, $M^\nu_\bD$ are defined as
ratio of partition functions (again, no degrees of freedom other than
the shape of the interface are considered):
$$ M^\nu_\bD = \frac{Z^\nu_{\bD}[\gamma]/Z^0_{\bD}[\gamma]}{
Z^\nu_{\bD}/Z^0_{\bD}}\  .$$
As in our previous examples, $M^\nu_\bD$ code for the off-critical
weights relative 
to the critical ones. On a finite lattice, they are typically well
defined but their existence 
in the continuum limit may be questioned.
This has to be analysed case by case.

Assume as in the SLE context that the interfaces emerge from the boundary
of $\bD$ so that the cut domains $\bD\setminus\gamma$ are also domains
of the complex plane.
The restricted partition function $Z_\bD[\gamma]$ is then proportional to the
partition function in the cut domain
$$Z_\bD^\nu[\gamma]= e^{{\cal E}^\nu_\bD(\gamma)} 
Z^\nu_{\bD\setminus\gamma} \textrm{ .}$$
The extra term ${\cal E}^\nu_\bD(\gamma)$ arises from the energy of the lattice
bonds which have been cut from $\bD$ to make $\bD\setminus\gamma$.
We call it the 'interface energy' of $\gamma$. It inherits from the
domain Markov property an additivity identity similar to the one
satisfied by the natural parameterization, i.e.
$${\cal E}^\nu_\bD(\gamma . \gamma')={\cal E}^\nu_\bD(\gamma)+ {\cal
  E}^\nu_{\bD\setminus \gamma}(\gamma')  \textrm{ .}$$
where $\gamma . \gamma'$ is the concatenation of successive segments
of the interface.

The off-critical weights then read:
\begin{eqnarray}
M^\nu_\bD(\gamma) = e^{ {\cal E}_\bD^\nu(\gamma)-{\cal E}_\bD^0(\gamma)}\
\frac{ Z^\nu_{\bD\setminus\gamma}/Z^0_{\bD\setminus\gamma} }{
Z^\nu_{\bD}/Z^0_{\bD}}.
\label{Mdiscrete}
\end{eqnarray}
This can be compared with (\ref{Msaw}). The presence of the energy term 
${\cal E}_\bD^\nu(\gamma)-{\cal E}_\bD^0(\gamma)$ in the continuum has also 
to be analysed case by case, see below. We furthermore point out that
there may be several natural choices of what to include in the
Boltzmann weights and different choices may lead to different
${\cal E}_\bD^\nu$ term\footnote{An easy example is the Ising model.
Suppose we have boundary conditions such that spins on the boundary
of the domain are fixed. Whether we include interactions of these
fixed spins with each other in our Hamiltonian and therefore in $Z_\bD$
obviously has a dramatic effect on the interface energy term while it
doesn't change the physics at all.} ---
to the extent that vanishing of this term can
be a question of convention.

To make contact with SLE, we also define a stochastic growth process
that describes the curve, in terms of which we define a filtration
on $\mathcal{C}$.
Consider portions of interfaces $\gamma[0,t]$, 
where the index $t$ specify say their lengths and
will be identified with the `time' of the process.
We may partition our configuration space according to these portions
at time $t$. The elements of the partition $\mathcal{Q}_t$ are denoted
by $\mathcal{C}_{\gamma_{[0,t]}}$, indexed by $\gamma[0,t]$ in such a
way that $c \in \mathcal{C}_{\gamma[0,t]}$ if and only if the
configuration $c$ gives rise to $\gamma[0,t]$ as a portion of the
interface.
Thus we have $\mathcal{C}= \bigcup_{\gamma[0,t]}\mathcal{C}_{\gamma[0,t]}$,
with $\mathcal{C}_{\gamma[0,t]}$ all disjoint.
By convention $\mathcal{Q}_0$ is the trivial partition with
the whole configuration space $\mathcal{C}$ as its single piece.
We assume these partitions to be finer as $t$ increases because
specifying longer and longer portions of interfaces defines
finer and finer partitions.
This  means that for any $s>t$ and any element $\mathcal{C}_{\gamma[0,t]}$
of the partition at time $t$ there exist elements of $\mathcal{Q}_s$
which form a partition of $\mathcal{C}_{\gamma[0,t]}$ (corresponding to
those $\gamma[0,s]$ which extend $\gamma[0,t]$). 
To any partition $\mathcal{Q}_t$ is
associated a $\sigma$-algebra $\sF^\gamma_t$ on $\mathcal{C}$, the one
generated by the elements of this partition.  Since these partitions
are finer as `time' $t$ increases, these constitute a filtration
$(\sF^\gamma_t)_{t \geq 0}$ on $\mathcal{C}$, i.e.
$\sF^\gamma_t \subset \sF^\gamma_s$ for $s>t$.
The fact that we trivially get a filtration simply means that
increasing `time' $t$
increases the knowledge on the system.
In the SLE context the information about the curve is encoded in the
driving process $(\xi_t)_{t \geq 0}$,
so this filtration $(\sF^\gamma_t)_{t \geq 0}$ becomes the
one generated by $\xi$.

On ${\cal C}$ with the filtration $(\sF^\gamma_t)_{t \geq 0}$, 
we may define two processes using either the critical $\PR^0$ or the
off-critical $\PR^\nu$ probability measures.
They differ by $M^\nu_\bD(\gamma[0,t])$ which can then be written as a 
conditioned expectation with respect to the critical measure
$$ M^\nu_\bD(\gamma[0,t]) = \frac{Z_\bD^0}{Z_\bD^\nu}\ 
\EX[\, \frac{w^\nu}{w^0} \,\vert \sF^\gamma_t\,]$$
Thus $M_\bD^\nu(\gamma[0,t])$ is a $P_\bD^0$-martingale and the two processes
differ by a martingale, which is the context in which Girsanov theorem applies.
It is similar to what we encountered in the SAW example. 
One of our aims is to (try to) understand how this tautological 
construction applies in the continuum.

\subsection{Continuum limit}

\subsubsection{Massive continuum limits in field theory}
In the continuum limit, the critical model should be described by a 
conformal field theory (CFT) and the critical measure on curves by SLE.
The Boltzmann weights are $e^{-S}$ with $S$ the action.
Off-critical perturbation is generated by a so-called perturbing field
$\Phi$ so that
$$ S = S^0 + \nu \int_\bD d^2z\, \Phi(z, \bar z)$$
with $S^0$ the conformal field theory action\footnote{For simplicity 
we assume that there is only one coupling 
constant and thus only one perturbing field.
Furthermore, renormalization properties of the field theory expression 
of the partition functions would need to be analysed. We shall not dive into 
this problem in view of so (low and) formal level we are at.}.
The ratio of the (off-critical) field theory partition function to the
CFT (critical) one is the expectation value
\begin{eqnarray}
\frac{\vev{\exp[-\nu\int_\bD d^2z\, \Phi(z,\bar z)] \; (\bcond)}_{\bD}}
     {\vev{(\bcond)}_{\bD}}
\label{zbyz}
\end{eqnarray}
where the brackets denote CFT expectations and
the boundary conditions $(\bcond)$ are implemented
by insertion of appropriate boundary operators, 
including in particular the operators that generate the interface. 

The coupling constant $\nu$ has 
dimension $2-h-\cc{h}$, linked to the scaling dimension $h+\cc{h}$ of
the perturbing operator $\Phi$. In our previous examples we determined
explicitly this dimension by looking at the way the scaling limit
is defined. We got (the perturbing operators in all cases are spinless,
$h=\cc{h}$):\\
(i) SAW: $\nu$ has dimension $d_\kappa=1+\kappa/8$, 
i.e. $d_{8/3}=4/3$ for $\kappa=8/3$ which is the value corresponding to
the SAW. 
The perturbing operator has dimension $h_\kappa+\cc{h}_\kappa=1-\kappa/8$, 
i.e. $h=\cc{h}=1/3$ for SAW. 
It is the operator $\Phi_{0;1}$ which is known to be the operator testing
for the
presence of the SLE curve in the neighbourhood of its point of insertion
(in particular its one-point function gives the probability for the SLE curve
to visit a tiny neighbourhood of a point in the complex plane,
\cite{BB-zig_zag}).
This had to be expected since the perturbation by the natural parametrization
as described in the first section just counts the number of lattice size boxes 
crossed by the curve. \\
(ii) LERW: the coupling constant has dimension $2$ so that the perturbing
operator has dimension $h=\cc{h}=0$ (up to logarithmic correction).
We shall identify it
either in terms of symplectic fermions or in terms of Brownian local time 
in the following sections.\\
(iii) Percolation: the coupling constant has dimension $3/4$ and therefore
the perturbing operator should have $h=\cc{h}=5/8$. This operator is the
bulk four-leg operator $\Phi_{0,2}$ testing for the presence of a
macroscopically pivotal point.

\subsubsection{Curves, RN-derivatives and interface energy}
\label{sec: interface energy}

Assuming (with possibly a posteriori justifications) that the
discrete martingale
(\ref{Mdiscrete}) has a nice continuum limit, one infers that
the off-critical measure $\EX^\nu[\,\cdots\,]$ 
and critical SLE measure $\EX[\,\cdots\,]$ on curves differ by a martingale 
(the Radon-Nikodym derivative exists) so that
$$ \EX^\nu[\, X \,] = \EX[\, M_t^\nu\, X \,] $$
for any $\sF^\gamma_t$-measurable observable with $M_t$ given by
the continuum limit of eq.(\ref{Mdiscrete}),
\begin{eqnarray} 
M_t^\nu= e^{ \Delta{\cal E}_\bD^\nu(\gamma_{[0,t]}) }\
\frac{ Z^\nu_{\bD\setminus\gamma_{[0,t]}} / Z^0_{\bD\setminus\gamma_{[0,t]}} }{
Z^\nu_{\bD}/Z^0_{\bD} }
\label{Mcontinuous}
\textrm{ .}
\end{eqnarray}
We expect the above ratio of partition functions to become the field
theory expression (\ref{zbyz}) in the continuum.
This is clearly a complicated (and useless) formula, but the existence of
$M_t^\nu$, at least in finite domain, is suggested by the physical intuition
that typical samples of the critical and off-critical interfaces look
locally similar on scales small compared to the correlation length
which is macroscopic.

As far as we know, there is no simple field theoretical formula for the surface
energy term $\Delta{\cal E}_\bD^\nu(\gamma_{[0,t]})$. However, to discuss
whether this term is present or not we may consider the discrete models
and propose criteria.

In the discrete setup we can typically write the offcritical Boltzmann
weight as $w^\nu = w^0 \, e^{-\sum_z \nu^{(a)}(z) \phi(z)}$, where
$\phi$ is a field by which we perturb the model. Under renormalization
it corresponds to the
scaling field $\Phi$ in the sense
that $a^{-h-\cc{h}} \phi(z)$ can in the limit $a \downarrow 0$ be replaced
by $\Phi(z)$. With our choice $\nu^{(a)}(z) = a^{2-h-\cc{h}} \nu(z)$,
the sum $\sum_{z \in \bD^{(a)}} \nu^{(a)}(z) \phi(z)$ becomes
$\int_\bD \nu(z) \Phi(z,\cc{z}) \ud^2 z$ in the continuum.
The martingale $M_t$ can be written in terms of
\debut \non
\EX_{\bD^{(a)}} \Big[ \exp \big( - \sum_{z \in \bD^{(a)}}
    \nu^{(a)}(z) \phi(z) \big) \Big| \sF^\gamma_t \Big]
\textrm{ .}
\fin
For example for SAW we have $\phi(z)=1$ if the walk passes through $z$
and $\phi(z)=0$ otherwise. For the Ising model in near critical
temperature, $\phi$ is the energy, most conveniently defined on
edges and not vertices, taking values $\pm 1$.

Let us assume, having in mind spin models with local interactions or SAW,
that $\phi(z)$ becomes determined ($\sF^\gamma_t$-measurable) for those $z$ that
are microscopically close to the curve $\gamma^{(a)}[0,t]$. Moreover we
must assume the domain Markov property. We then get
\debut \non
M_t^{(a)} & \; = \; & \const \; \times \;
    \exp \big( - \sum_{z \in \gamma^{(a)}[0,t]}
    a^{2-h-\cc{h}} \nu(z) \phi(z) \big) \\
\non
& & \qquad \times \;
    \EX_{\bD^{(a)}\setminus \gamma^{(a)}[0,t]} \Big[
    \exp \big( \sum_{z \in \bD^{(a)} \setminus \gamma^{(a)}[0,t]}
    a^{2-h-\cc{h}} \nu(z) \phi(z) \big) \Big]
\textrm{ .}
\fin
The first part corresponds to the ``interface energy'' and the latter
to the same model in the remaining domain. Since the number of points
microscopically close to the interface is of order $ \sim a^{-d}$
(where $d$ is the fractal dimension of the curve) and 
$\phi$ is typically bounded, the interface energy term should vanish
in the continuum if $2-d-h-\cc{h}>0$. In the case $2-d-h-\cc{h}=0$ there
may remain a finite interface energy in the continuum.
If $2-d-h-\cc{h} < 0$ some additional cancellations would have to
take place if the expressions were to have continuum limits.

In view of the above, we notice that for example for percolation
$h=\cc{h}=5/8$ and $d=7/4$, so we must be careful.
Indeed, the near critical percolation interfaces have been considered
in  \cite{NW-offcritical_percolation} and they have been shown not to
be absolutely continuous with respect to the critical ones.
The SAW is just the marginal case: $h=\cc{h}=1/3$
and $d=4/3$ so that $2-d-h-\cc{h}=0$. Indeed we expect a finite
interface energy term $L_\bD(\cdot)$ in the continuum.

However, the LERW doesn't quite fit into the above setup as such ---
some long range interactions are present.
The field $\phi(z)$ is now the number of visits of the underlying
walk to $z$, denoted by $\ell_{\bD^{(a)}}(z)$ (for a more formal
definition, see section \ref{clL}). It splits to
$\ell_\bD(z) = \ell^{(t)}_{\bD}(z) + \ell_{\bD \setminus \gamma[0,t]}(z)$
where the former represents visits of the walk to $z$ until the last time
it comes to $\gamma_t$ and the latter represents the visits to $z$ of
the walk after this time. The quantity $\ell^{(t)}_\bD$ is not
$\sF^\gamma_t$-measurable, but conditional on $\sF^\gamma_t$ it is independent of
the walk after the last time it came to $\gamma_t$, see e.g.
\cite{LSW-LERW_and_UST}. Thus we have
\debut \non
M_t^{(a)} & \; = \; & \const \; \times \;
    \EX \big[ \exp \big( - a^2 \sum_{z \in \bD^{(a)}}
    \nu(z) \ell^{(t)}_{\bD^{(a)}}(z) \big) \big| \sF^\gamma_t \big] \\
\non
& & \qquad \times \;
    \EX_{\bD^{(a)}\setminus \gamma^{(a)}[0,t]} \Big[
    \exp \big( a^2 \sum_{z \in \bD^{(a)} \setminus \gamma^{(a)}[0,t]}
    \nu(z) \ell_{\bD^{(a)}\setminus \gamma^{(a)}[0,t]}(z) \big) \Big]
\textrm{ .}
\fin
The former term is again a property of the curve $\gamma[0,t]$ and the
domain: it can be written in terms of random walk bubbles along the
curve. The bubbles may occasionally reach far away and thus they feel
the values of $\nu$ in the whole domain. In this sense an interface
energy is present in the LERW (with our conventions).
The crucial difference is, however, that
sites microscopically close to the curve don't contribute to the
continuum limit. The values of $\ell^{(t)}_{\bD^{(a)}}$ on the curve
remain of constant order (or diverges logarithmically
still in accordance with $h=\cc{h}=0$) while the number of sites
close to the curve is $\sim a^{-d}$ with $d=5/4$. The contribution
along the curve to the interface energy thus vanishes like $\sim a^{3/4}$.
We will use repeatedly the possibility to change between domains
$\bD$ and $\bD \setminus \gamma[0,t]$ in integrals of type
$\int \nu(z) \ell(z) \ud^2 z$.

\subsubsection{Field theoretic considerations of the RN-derivative}

If it were correct to use the field theory expression (\ref{zbyz}) in
formula (\ref{Mcontinuous}) in the continuum limit without an
interface energy term, we would have to first order in $\nu$
\debut \label{eq: M from QFT}
M_t^\nu= 1 +
    \nu[\, \int_{\bD\setminus\gamma_{[0,t]} } \ud^2z\, N_t(z) 
    -\int_{\bD} d^2z\, N_0(z) \,] +\cdots
\fin
with
\begin{eqnarray}
N_t(z) := \frac{ \vev{\Phi(z,\bar z) \; (\bcond) }_{\bD \setminus \gamma_{[0,t]}}}
{\vev{(\bcond)}_{\bD\setminus\gamma_{[0,t]}} }
\label{Fpert}
\end{eqnarray}
Here the $(\bcond)$ refers to insertion of the appropriate boundary
operators.
For any point $z$, this ratio $N_t(z)$ of correlation functions is 
a SLE (local) martingale, see e.g.
\cite{BBH-dipolar_SLE, BBK-multiple_SLEs} and discussion in section
\ref{sec: CFT martingales}.
This is a good sign since $M_t$, if it exists, should be a martingale
by construction.
In the case of LERW, we will see also in section \ref{sec: CFT martingales}
that $N_t(z)$ thus defined is a sum of two parts, precisely corresponding
to $\ell_{\bD\setminus\gamma[0,t]}$ and $\ell_\bD^{(t)}$, and $N_t(z)$
will indeed be closely related to the Radon-Nikodym derivative $M_t$.

\subsection{Off-critical drift term and Girsanov theorem}
\label{sec: Girsanov}

As argued above, the off-critical expectations are related to the 
critical ones by insertion of the martingale $M_t^\nu$:
$$ \EX^\nu[\, X \,] = \EX[\, M_t^\nu\, X \,] $$
for any $\sF^\gamma_t$-measurable observable.
With some regularity assumptions, this is a situation in which
one may apply the Girsanov's theorem, which relates the
decompositions of semimartingales in two probability measures one of
which is absolutely continuous with respect to the other. A simple
illustration of the idea of Girsanov's theorem is given in appendix
\ref{app: Girsanov}.

By definition of chordal SLE, $\xi_t=\sqrt{\kappa} B_t$ with $B_t$ a
Brownian with respect to the critical measure $\PR^0$. Since $M_t$
is a martingale, its It\^o derivative is of the form 
$M_t^{-1} \, \ud M_t = V_t \, \ud B_t$. Girsanov theorem tells
us we may write
$$ \ud \xi_t \; = \;
\sqrt{\kappa} \; \ud B_t' + \sqrt{\kappa} \,V_t \; \ud t $$
where $B_t'$ is Brownian motion with respect to the off-critical measure
$\PR^\nu$.

In other words, weighting the expectation by the martingale $M_t$ 
adds a drift term to the stochastic evolution of the driving process
$\xi_t$.

In the present context the martingale, if it exists, is given by 
(\ref{Mcontinuous}) and it seems hopeless to compute and use 
the drift term directly. However, if the field theoretic expression
(\ref{eq: M from QFT}) is correct and we may omit contributions along
the curve, we have to first order in perturbation simply
\debut \non
V_t \, \ud B_t \; = \; 
\nu \int_{\bD \setminus \gamma[0,t]} \ud^2 z \,
\big( \ud \, N_t(z) \big)
\textrm{ .}
\fin
In this situation we'd have under $\PR^\nu$ the following drift,
to first order in perturbation $\nu$
\debut \non
\ud \xi_t \approx \sqrt{\kappa} \; \ud B_t' + \sqrt{\kappa} \nu \;
    \int_\bD \ud^2 z \; \big( \ud \qv{B,N}_t \big)
\textrm{ ,}
\fin
where $\qv{B,N}_t$ is the quadratic covariation of $B$ and $N$,
$\ud \qv{B,N}_t = V_t \; \ud t$.
In section \ref{sec: perturbed SLEs} we will argue in two different
ways that the above formula applies to the LERW case. The explicit
knowledge of $N_t(z)$ will of course make this more concrete.

The same change of drift applies to variants of SLE, where the
driving process contains a drift to start with. If a process
has increments $\ud \xi_t = \beta \; \ud B_t + \alpha \; \ud t$,
it is only the random part of the increment $\beta \; \ud B_t$
that is affected by the change of probability measure: the
deterministic increments remain otherwise unchanged, but they
gain the additional term discussed above from the change of the
random one.

\section{Critical and off-critical LERW}
\label{sec: LERW}

In our attempt to gain insight to curves out of
the critical point we now concentrate on the concrete example
of loop-erased random walks (LERW).
It is worth noticing that the powerful method of
Schramm-Loewner evolutions (SLE) that applies very generally
to critical (conformally invariant) statistical mechanics in
two dimensions, was in fact first introduced with an application to LERW
\cite{Schramm-LERW_and_UST}. And one of the major early successes
of SLEs was indeed the proof that scaling limit of (radial)
LERW is (radial) SLE${}_2$
\cite{LSW-LERW_and_UST}. We will not consider the radial LERW,
but very natural variants of the same idea, namely chordal and dipolar
LERW: in chordal setup the curves go from a boundary point
${x_0} \in \bdry \bD$ to another boundary point ${x_\infty} \in \bdry \bD$,
and in the dipolar setup from
a boundary point ${x_0}$ to a boundary arch $S \subset \bdry \bD$.
Scaling limits of these and other LERW variants at criticality have
been studied mathematically in \cite{Zhan-scaling_limits_of_LERW}.

\subsection{Continuum limit of LERWs} \label{clL}
The discrete setting for LERWs was described in the introduction
and we gave a formula for the offcritical measure in terms of the
random walks: the relative weight was
\debut
e^{- \sum_{0 \leq j \leq \tau^{\RW}_\bD} a^2 \nu(W_j)}
\textrm{ .}
\fin
There's an alternative way of writing
the Boltzmann weights of the walks on lattice $\bD^{(a)}$ of mesh $a$.
Let
$\ell^{(a)}(z) = \# \{ 0 \leq j < \tau^{\RW}_{\bD^{(a)}} :
W^{(a)}_j = z \}$
be the number of visits to $z \in \bD^{(a)}$ by the walk $W^{(a)}$.
Then the Boltzmann weight is
\debut \non
\prod_{0 \leq j < \tau^{{\rm rw}}_{\bD^{(a)}}} \mu^{(a)}(W_j) & = &
    \prod_{z \in \bD^{(a)}} \mu^{(a)}(z)^{\ell(z)}
\textrm{ .}
\fin
In terms of the $\nu(z) = a^{-2} \log (\mu_c/\mu^{(a)}(z))$ we can write the
partition function as an expected value for a random walk $W^{(a)}$
started from $w^{(a)} \in \bD^{(a)}$
\debut
\label{eq: discrete partition function 2}
Z^{\bD^{(a)} ; w^{(a)} ; S^{(a)}}_\nu & = & \EX^{w^{(a)}}_{\RW}
    \Big[ \exp \big( - \sum_{z \in \bD^{(a)}}
    a^2 \nu(z) \ell^{(a)}(z) \big) \;
    \ind_{W^{(a)}_{\tau^{\RW}_{\bD^{(a)}}} \in S} \Big]
\textrm{ .}
\fin

\bigskip

We will take the continuum limit by letting the lattice spacing $a$ tend
to zero and choosing $\bD^{(a)}$ that approximate a given
open, simply connected domain $\bD$. The starting points $w^{(a)}$
approximate $w \in \bD$ and the target set $S^{(a)}$ approximate
$S \subset \bdry \bD$.
Simple random walks $W^{(a)}$ on the lattice should be scaled according
to $B^{(a)}_t = W^{(a)}_{\lfloor t/a^2 \rfloor}$, so that $B_t^{(a)}$
converges to two-dimensional Brownian motion $\Bplan_t$. In the limit
$a^2 \sum_{z \in \bD^{(a)}}$ becomes an integral $\int_{\bD} \ud^2 z$
and the partition function
(\ref{eq: discrete partition function 2}) becomes
\debut
\label{eq: continuum partition function 0}
Z^{\bD; w ; S}_\nu & = &
\EX^{w}_{\BM} \Big[ \exp \Big( -\int_D \ud^2 z \; \nu(z) \ell(z) \Big)
    \; \ind_{{\Bplan}_{\tau_D} \in S} \Big]
\textrm{ ,}
\fin
where $\ell(z)$ needed no rescaling: it is the limit of $\ell(z^{(a)})$
with $z^{(a)} \in \bD^{(a)}$ approximating $z \in \bD$.
This way $\ell (z)$ becomes the Brownian local time:
it has an interpretation as the
occupation time density
\debut \non
\int_0^{\tau_\bD} F({\Bplan}_t) \ud t = \int_{\bD} \ell(z) F(z) \; \ud^2 z
\textrm{ ,}
\fin
a discete analogue of which we already used for $F = \nu$ to obtain
the alternative expression for the Boltzmann weights.

By comparing (\ref{eq: continuum partition function 0}) with (\ref{zbyz}),
we see that the Brownian local time $\ell(z)$, although not a CFT operator,
plays a role very analogous to the perturbation $\Phi$. Similarly, 
``$\ind_{\textrm{exit in $S$}}$'' together with ``start from $w$''
impose the boundary conditions.

\bigskip

\emph{Remark:}
Our notation $\ell(z)$ is not totally fair, but in line with other
traditional field theory notation. It would be more appropriate
to consider $\ell$ as a random positive Borel measure on $\bD$ with
finite positive total mass $\tau_\bD$.
This measure is supported on the graph
${\Bplan}[0,\tau_\bD) \subset \bD$ of the Brownian
motion, which has Lebesgue measure $0$ (although its Hausdorff dimension
is $2$). Therefore $\ell$ can not be absolutely continuous w.r.t.
Lebesgue measure, as our notation suggests: we'd like $\ell(z)$ to be
defined pointwise as the density of the occupation time
$\ell$ with respect to Lebesgue measure.
However, as usual in field theory, it is possible to make sense of
pointwise correlation functions as long as the insertions are not at
coinciding points and we will stick to the convenient notation $\ell(z)$
although it seems to misleadingly suggest a pointwise definition of $\ell$.

\subsubsection{Continuum partition functions in the half-plane}
\label{sec: partition functions}
We now choose as our domain the upper half plane
$\bH = \{ z \in \bC : \im z > 0 \}$ and as the target set an
interval $S = [x_+,x_-]$.
The partition function
(\ref{eq: continuum partition function 0}) can be written in terms of
a Brownian expectation value 
\debut \non
Z^{w ; [x_+, x_-]}_\nu & = &
\EX^{w}_{\BM} \Big[ \exp \Big(-\int_0^{\tau_\bH} \nu({\Bplan}_s) \, \ud s \Big) \;
    \ind_{{\Bplan}_{\tau_\bH} \in [x_+,x_-]} \Big] \\
\non
& = &
\EX^{w}_{\BM} \Big[ \exp \Big( -\int_\bH \ud^2 z \; \nu(z) \ell(z) \Big)
    \; \ind_{{\Bplan}_{\tau_\bH} \in [x_+,x_-]} \Big]
\fin
with $\tau_\bH = \inf \{ t \geq 0 : {\Bplan}_t \notin \bH\}$ the exit time
from the half-plane.

We would like to let the LERW start from the boundary, that is
take $z \rightarrow x_0 \in \bdry \bD$.
In the limit $z \rightarrow {x_0}$ the partition function vanishes like
$Z^{{x_0} + i \delta ; [x_+, x_-]}_\nu \sim \delta \times (\cdots)$
so to obtain a nontrivial limit, we set
\debut
\label{eq: dipolar partition function 1}
Z^{{x_0}; [x_+, x_-]}_\nu & = & \lim_{\delta \rightarrow 0} \frac{1}{\delta}
    \EX^{{x_0} + i \delta}_{\BM}
    \Big[ \exp \Big(-\int_0^{\tau_\bH} \nu({\Bplan}_s) \ud s \Big) \;
    \ind_{{\Bplan}_{\tau_\bH} \in [x_+,x_-]} \Big] \\
\label{eq: dipolar partition function 2}
& = & \lim_{\delta \rightarrow 0} \frac{1}{\delta}
    \EX^{{x_0} + i \delta}_{\BM}
    \Big[ \exp \Big( -\int_\bH \ud^2 z \; \nu(z) \ell(z) \Big)
    \; \ind_{{\Bplan}_{\tau_\bH} \in [x_+,x_-]} \Big]
\textrm{ .}
\fin
Furthermore, we may wish to shrink the target set $S = [x_+,x_-]$ to
a point ${x_\infty}$ so as to obtain a chordal LERW, nontrivial limit is
obtained if we set
\debut \non
Z^{{x_0}; {x_\infty}}_\nu & = & \lim_{\delta, \delta' \rightarrow 0}
    \frac{1}{\delta \, \delta'}
    \EX^{{x_0} + i \delta}_{\BM}
    \Big[ \exp \Big(-\int_0^{\tau_\bH} \nu({\Bplan}_s) \ud s \Big) \;
    \ind_{{\Bplan}_{\tau_\bH} \in [{x_\infty} - \delta', {x_\infty} + \delta']} \Big]
\textrm{ .}
\fin

In the unperturbed case $\nu=0$, we have $Z^{{x_0}; [x_+,x_-]}_0
= \frac{1}{\pi} \big( \frac{1}{{x_0}-x_-} - \frac{1}{{x_0}-x_+} \big)
= \frac{1}{\pi} \frac{x_- - x_+}{({x_0}-x_-) ({x_0}-x_+)}$
and $Z^{{x_0}; {x_\infty}}_0 = \frac{2}{\pi} ({x_\infty}-{x_0})^{-2}$.
The former is indeed the partition function of a dipolar SLE${}_2$
and the latter is that of chordal SLE${}_2$ from ${x_0}$ to ${x_\infty}$,
see \cite{BBH-dipolar_SLE, BBK-multiple_SLEs, Kytola-SLE_kappa_rho,
Kytola-local_mgales}.

Partition functions with a nonzero perturbation will be considered
in more detail in section \ref{sec: subinterval hitting}.

\subsubsection{The perturbation and conformal transformations}
\label{sec: conformal maps of local time}
The perturbation $\ell(z)$ corresponds to an operator of dimension
zero. According to a general argument that can be found e.g. in
\cite{Cardy-scaling_and_renormalization}, this fact already
manifested itself when we observed that
no rescaling under renormalization was needed in its continuum
definition,
$a^{\Delta_\ell} \, \ell^{(a)}(v^{(a)}) \rightarrow \ell(z)$ with
$\Delta_\ell = 0$.
From its definition as a local time of 2-d Brownian motion
we can also directly check how $\ell(z)$ transforms under conformal
transformations. The local time $\ell(z)$ gives us the occupation time
in the following sense: if $F: \bD \rightarrow \bR$, then
\debut \non
\int_0^{\tau_\bD} F({\Bplan}_t) \; \ud t = \int_\bD F(z) \ell(z) \; \ud^2 z
\textrm{ .}
\fin
Taking in place of $F$ an approximate delta function, we see that
$\ell(z) = \int_0^{\tau_\bD} \delta({\Bplan}_t - z) \, \ud t$.

Let $f: \bD \rightarrow \tilde{\bD}$ be conformal and
$({\Bplan}_t)_{t \in [0,\tau_\bD]}$ Brownian motion in $\bD$ started from
$w \in \bD$ and stopped upon exiting the domain
$\tau_\bD = \inf \{ t \geq 0 : {\Bplan}_t \notin \bD \}$. Then a direct
application of Ito's formula tells us that
$(f({\Bplan}_t))_{t \in [0,\tau_\bD]}$ is a (two-component) martingale in
$\tilde{\bD}$, started from $\tilde{w} = f(w)$, and the quadratic
variation of its components is
$\ud \qv{F({\Bplan})_j, F({\Bplan})_k}_t \; = \;
\delta_{j,k} |f'({\Bplan}_t)|^{2} \, \ud t$, ($j,k=1,2$).
The time changed
process $\tilde{\Bplan}_s = f({\Bplan}_{t(s)})$ with 
$\ud s = |f'({\Bplan}_t)|^{2} \, \ud t$ is a Brownian motion in $\tilde{\bD}$,
started from $\tilde{w} = f(w)$.

Given $\tilde{F}: \tilde{\bD} \rightarrow \bR$ we set
$F = \tilde{F} \circ f : \bD \rightarrow \bR$
and we have by definitions
\debut \non
\int_{\tilde{\bD}} \tilde{F}(z') \ell_{\tilde{\bD};w'}(z') \; \ud^2 z'
& = & \int_0^{\tilde{\tau}_{\tilde{\bD}}} \tilde{F}(\tilde{\Bplan}_s) \; \ud s \\
\non
& = & \int_0^{\tau_\bD} F(f({\Bplan}_t)) |f'({\Bplan}_t)|^{2} \; \ud t \\
\non
& = & \int_\bD F(z) |f'(z)|^{2} \ell_{\bD;w}(z) \; \ud^2 z
\fin
If $\tilde{F}$ is an approximate delta function at $z'=f(z)$, then
$|f'|^2 \times F$ is an approximate delta at $z$ and we conclude
that $\ell$ indeed transforms as a scalar
\debut \non
\ell_{f(\bD);f(w)}(f(z)) 
\; \overset{\textrm{in law}}{=} \;
\ell_{\bD;w}(z)
\textrm{ .}
\fin

We will later in section \ref{sec: CFT perturbing operator}
identify the conformal field theory equivalent of
$\ell(z)$ and exhibit its corresponding transformation properties.

\subsubsection{Brownian local time expectations}
\label{sec: local time correlations}
The multipoint correlation functions of the perturbing operator
$\ell(z)$ are the basic building blocks of the perturbative analysis
of LERW near critical point since we can expand the partition function
(\ref{eq: dipolar partition function 2}) in powers of the small
perturbation $\nu = \eps \tilde{\nu}$
\debut \non
Z^{{x_0}; [x_+,x^-]}_{\eps \tilde{\nu}} & = &
    \lim_{\delta \rightarrow 0} \frac{1}{\delta}
    \EX^{{x_0} + i \delta}_{\BM}
    \Big[ e^{-\eps \int_\bH \tilde{\nu}(z) \ell (z) \; \ud^2 z}
    \; \ind_{{\Bplan}_{\tau_D} \in [x_+,x_-]} \Big] \\
\non
& = & Z^{{x_0}; [x_+,x_-]}_0 + \sum_{n=1}^\infty \frac{\eps^{n}}{n!} \mint 
    \ud^2 z_1 \! \cdots \! \ud^2 z_n \;
    \tilde{\nu}(z_1) \cdots \tilde{\nu}(z_n) \\
\label{eq: dipolar perturbative expansion}
& & \qquad \times \Big( \lim_{\delta \rightarrow 0} \frac{1}{\delta}
    \EX^{{x_0} + i \delta}_{\BM}
    \Big[ \ell(z_1) \cdots \ell(z_n)
    \; \ind_{{\Bplan}_{\tau_D} \in [x_+,x_-]} \Big] \Big)
\textrm{ .}
\fin
Next we will compute these explicitly and afterwards we'll find the
field theoretic interpretation.

For a smooth compactly supported function $f: \bD \rightarrow \bR$,
let $\ell_{f} = \int_0^{\tau_\bD} f({\Bplan}_t) \; \ud t$. Consider the
correlation function
\debut \non
C^S_{f_1, \ldots, f_n} (w) & = & \EX^{w}_\BM \Big[
    \Big( \prod_{j=1}^n \ell_{f_j} \big)
    \; \ind_{{\Bplan}_{\tau_\bD} \in S} \Big]
\textrm{ .}
\fin
If $\sigma \leq \tau_\bD$ is
a stopping time of the Brownian motion, then write
$\ell_f = \int_0^\sigma f({\Bplan}_t) \; \ud t +
\int_\sigma^{\tau_\bD} f({\Bplan}_t) \; \ud t = \ell^{\leq \sigma}_f +
\ell^{>\sigma}_f$. The part $\ell^{\leq \sigma}_f$ is
$\sF^{\BM}_\sigma$-measurable while $\ell^{> \sigma}_f$ depends
on $\sF^{\BM}_\sigma$ only through ${\Bplan}_\sigma$. Obviously
we have $\ell^{\leq 0}_f = 0$ and $\ud \ell^{\leq t \minim \tau_\bD}_f =
\ind_{t \leq \tau_\bD} f({\Bplan}_t) \; \ud t$.
By the strong Markov property we have
\debut \non
& & \EX_\BM^w \Big[ \Big( \prod_{j = 1}^n \ell_{f_j} \Big) \;
    \ind_{{\Bplan}_{\tau_\bD} \in S}  \Big| \sF^\BM_{t \minim \tau_\bD} \Big] \\
\non
& = & \sum_{J \subset \{ 1,\ldots,n \}}
    \Big( \prod_{j \in J} \ell^{\leq t \minim \tau_D}_{f_j} \Big)
    \; \times \; C^S_{(f_j)_{j \in \complement J}}({\Bplan}_{t \minim \tau_D})
\fin
and this is a martingale by construction. It is also a continuous
semimartingale and its It\^o drift
\debut \non
& & \sum_{J \subset \{ 1,\ldots,n \}} \Big\{
    \sum_{k \in J} \ind_{t \leq \tau_\bD} f_k({\Bplan}_t) \; \Big(
    \prod_{j \in J\setminus \{k\}} \ell^{\leq t \minim \tau_\bD}_{f_j} \Big)
    \; \times \; C^S_{(f_j)_{j \in \complement J}}({\Bplan}_{t \minim \tau_\bD}) \\
\non
& & \qquad + \Big( \prod_{j \in J} \ell^{\leq t \minim \tau_\bD}_{f_j} \Big)
    \; \times \; \frac{1}{2} \ind_{t \leq \tau_\bD} \; \lapl
    C^S_{(f_j)_{j \in \complement J}}({\Bplan}_{t \minim \tau_\bD})
    \Big\}
\fin
should vanish. At $t=0$ we have simplifications due to
$\ell^{\leq 0}_f = 0$ and $\ind_{0 \leq \tau_\bD} = 1$, so this reduces
to a useful differential equation for $C^S_{f_1, \ldots, f_n}$
\debut \non
\frac{1}{2} \lapl C^S_{f_1, \ldots, f_n} (w)
    + \sum_{k=1}^n f_k(w) C^S_{(f_j)_{j \neq k}} (w) \; = \; 0
\fin
in terms of correlation functions of type $C^S_{f_1, \ldots, f_{n-1}}$.
Boundary conditions for $n \geq 1$ are zero, and for $n=0$ case the
correlation function is just the harmonic measure of $S$,
$C^S_\emptyset(w) = H_\bD(w;S)$.

We are interested in replacing $f_j(z)$ by $\delta(z-z_j)$,
in which case we denote the correlation function by
$C^S(w;z_1, \ldots, z_n)$. 
It is then straightforward to solve the recursion and the result is
\debut \non
C^S(w;z_1, \ldots, z_n) & = &
    \EX_\BM^w \Big[ \ell(z_1) \cdots \ell(z_n) \;
    \ind_{{\Bplan}_{\tau_\bD} \in S} \Big] \\
\non
& = & \sum_{\pi \in S_n} G_\bD(w,z_{\pi(1)}) \; \Big( \prod_{j = 2}^n
    G_\bD (z_{\pi(j-1)}, z_{\pi(j)}) \Big) \; H_\bD(z_{\pi(n)},S)
\textrm{ ,}
\fin
where $G_\bD$ is the Green's function
$\lapl_z G_\bD(z,w) = -2 \delta(z-w)$
with Dirichlet boundary conditions $G_\bD(z,w) \rightarrow 0$ as
$z \rightarrow \bdry \bD$.
To get the multipoint correlation function for Brownian motion
conditioned to exit through $S$, we must divide by
$\EX^w_\BM [\ind_{{\Bplan}_{\tau_\bD} \in S}] = H_\bD(w, S)$, which we remind
is also the partition function at criticality. The ratio has
a nontrivial limit as we take $w$ to the boundary of the domain.
Alternatively, we can regularize both the correlation function and the
partition function in the same manner, as suggested also by formula
(\ref{eq: dipolar perturbative expansion}). In the half-plane $\bH$
with $S = [x_+,x_-]$, regularized as in section
\ref{sec: partition functions} we have
\debut \non
& & C^{{x_0} ; [x_+,x_-]} (z_1, \ldots, z_n) \\
\non
& := & \lim_{\delta \rightarrow 0} \frac{1}{\delta}
    \EX_\BM^{{x_0} + i\delta} \Big[ \ell(z_1) \cdots \ell(z_n)
    \; \ind_{{\Bplan}_{\tau_D} \in [x_+,x_-]} \Big] \\
\label{eq: multipoint function dipolar}
& = & \sum_{\pi \in S_n} K_\bH ({x_0}, z_{\pi(1)}) \; \Big( \prod_{j = 2}^n
    G_\bH (z_{\pi(j-1)}, z_{\pi(j)}) \Big) \; H_\bH (z_{\pi(n)};[x_+,x_-])
\textrm{ ,}
\fin
\begin{figure}
\includegraphics[width=1.0\textwidth]{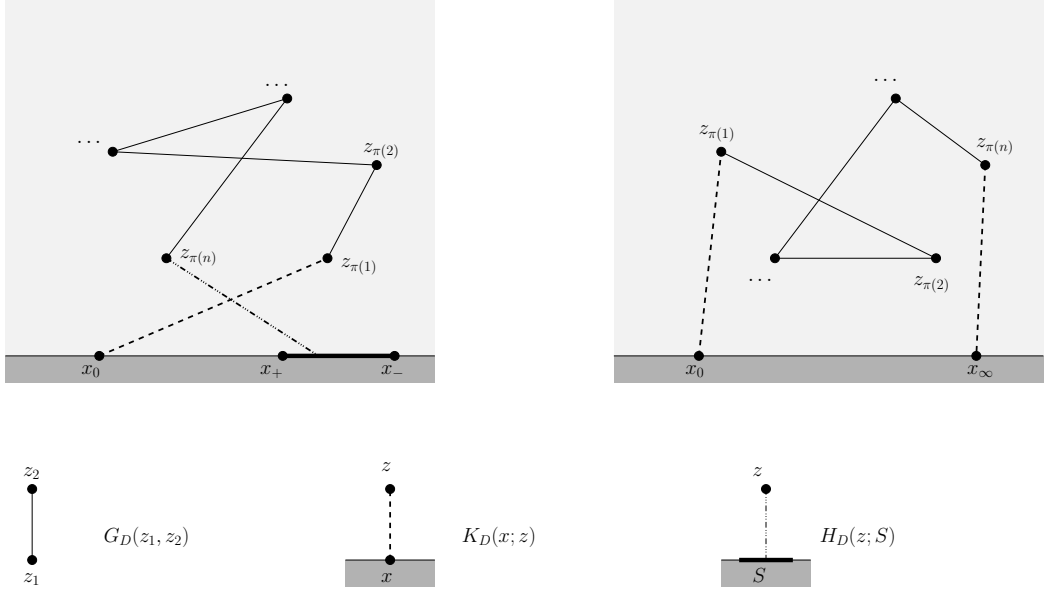}
\caption{\emph{Example diagrams representing the terms in the local time
multipoint correlation functions.
Dipolar case (\ref{eq: multipoint function dipolar}) is on the 
left and chordal case (\ref{eq: multipoint function chordal})
on the right.}}
\label{fig: local time diagram}
\end{figure}
with explicit expressions
\debut \non
G_\bH (z,w) & = & -\frac{1}{\pi} \log \big| \frac{z-w}{z-\cc{w}} \big| \\
\non
K_\bH ({x_0}, z) & = & -\frac{2}{\pi} \; \im \big( \frac{1}{z-{x_0}} \big) \\
\non
H_{\bH}(z,[x_+,x_-]) & = & \frac{1}{\pi} \; \im \Big(
    \log \frac{z-x_-}{z-x_+} \Big)
\textrm{ .}
\fin
It is convenient to represent the terms in this result diagrammatically
as in figure \ref{fig: local time diagram}.
The chordal case is obtained by limit $\delta' \rightarrow 0$ with
choice $x_\pm = {x_\infty} \mp \delta'$,
\debut \non
& & C^{{x_0} ; {x_\infty}} (z_1, \ldots, z_n) \\
\non
& := & \lim_{\delta \rightarrow 0} \frac{1}{\delta \, \delta'}
    \EX_\BM^{{x_0} + i\delta} \Big[ \ell(z_1) \cdots \ell(z_n)
    \; \ind_{{\Bplan}_{\tau_D} \in [{x_\infty}-\delta', {x_\infty}+\delta']} \Big] \\
\label{eq: multipoint function chordal}
& = & \sum_{\pi \in S_n} K_\bH ({x_0}, z_{\pi(1)}) \; \Big( \prod_{j = 2}^n
    G_\bH (z_{\pi(j-1)}, z_{\pi(j)}) \Big) \; K_\bH ({x_\infty}, z_{\pi(n)})
\textrm{ .}
\fin

\subsection{On conformal field theory of LERWs}

It is known from general arguments that SLE${}_\kappa$ corresponds
to conformal field theory of central charge
$c = \frac{(6-\kappa)(3\kappa-8)}{2 \kappa}$,
\cite{BB-CFTs_of_SLEs}, so that LERWs should have $c=-2$.
But we can be more specific about the CFT appropriate for our case.

First of all, LERWs are ``dual'' to uniform spanning trees (UST)
\cite{Wilson-generating_UST, Schramm-LERW_and_UST, LSW-LERW_and_UST},
for which fermionic field theories have been given
\cite{CJHSS-fermionic_trees_and_forests}, see also
\cite{Majumdar-exact_fractal_dimensions_of_LERW}.
Indeed a field theory of free symplectic fermions would have central charge
$c=-2$, \cite{Kausch-symplectic_fermions}. The theory is Gaussian.
It has two basic fields
$\chi^+$ and $\chi^-$ whose correlation functions in domain $\bD$
(Dirichlet boundary conditions) are determined by
\debut \non
\vev{\chi^\alpha (z,\cc{z}) \chi^\beta (w,\cc{w})}
    = J^{\alpha \beta} G_\bD(z,w)
\textrm{ ,}
\fin
with $J^{+ +}=0=J^{- -}$, $J^{+ -} = 1 = - J^{- +}$,
and the Wick's formula.

The fields $\chi^\alpha (z,\cc{z})$ are fermionic but scalars, meaning that
they transform like scalars under conformal transformations.
We shall also be interested in the composite operator $ \no{\chi^-\chi^+}$
which has to be defined via a point splitting to remove the short
distance singularity 
$$\; \no{ \chi^- \, \chi^+ } (z,\cc{z})
=\lim_{z\to w} \; \chi^-(z,\cc{z})\,\chi^+(w,\cc{w}) 
-\frac{1}{2\pi}\log|z-w|^2
$$
Due to this regularisation, $\no{\chi^-\chi^+}$ transforms with a logarithmic
anomaly under conformal transformations:
\begin{equation}\label{eq:loganom}
\no{ \chi^- \, \chi^+ } (z,\cc{z}) \to \no{ \chi^- \, \chi^+ } (g(z),\cc{g(z)})
- \frac{1}{2\pi}\log|g'(z)|^2 \textrm{ .} \end{equation}

The stress tensor is
$T(z) = 2 \pi \; \no{ \partial_z \chi^+(z,\cc{z})\, \partial_z \chi^-(z,\cc{z}) }$ 
with the normal ordering defined by a point splitting similar as above. 
It is easy to verify that both operators $\partial_z\chi^\alpha$ are operators 
of dimension $1$ satisfying the level two null vector equation 
$(L_{-2}-1/2L_{-1}^2)\partial_z\chi^\alpha=0$ with $L_n$,
$T(z)=\sum_n L_n z^{-n-2}$, the Virasoro generators.
It is this equation which helps identifying the symplectic fermions
as the CFT associated to LERW.
We will be able to identify some other fields with natural LERW quantities, 
although there are some important ones for which a good understanding 
is still lacking (to us).

\subsubsection{Boundary changing operators}
\label{sec: boundary operators}
The partition functions without perturbation involve only boundary
operators that account for the LERW starting from ${x_0} \in \bR$ and
aiming at $S \subset \bR$. We will identify them below.

We consider the symplectic fermion field theory in the upper half
plane $\bH$. Let us define the boundary fields $\psi^\pm$ as normal
derivatives of $\chi^\pm$ on the real axis
\debut \non
\psi^\pm ({x_0}) = \lim_{\delta \rightarrow 0} \frac{1}{\delta}
\chi^\pm({x_0}+i\delta,{x_0}-i\delta)
\textrm{ .}
\fin
The level two null field equation says the fields $\psi^+$ and
$\psi^-$ can account for starting point and end point of SLE${}_2$
curves \cite{BB-CFTs_of_SLEs, BBH-dipolar_SLE, BBK-multiple_SLEs}
(see also section \ref{sec: CFT martingales}).
And indeed, the two point function
$\vev{\psi^+({x_0}) \psi^-({x_\infty})} = \frac{2}{\pi} ({x_\infty} - {x_0})^{-2}$
reproduces our partition function in the chordal setup,
compare with sections \ref{sec: SLE basics} and
\ref{sec: partition functions}.

Let us then remark that the
dipolar LERW from ${x_0}$ to $[x_+,x_-]$, conditioned to hit a point
${x_\infty} \in [x_+,x_-]$ is just the chordal LERW from ${x_0}$ to ${x_\infty}$
as follows directly from the definitions.
It has been pointed out in \cite{BBKen} that
$\kappa = 2$ is the only value for which the corresponding property
holds for dipolar and chordal SLE${}_\kappa$.

Following the above remark, we decompose the dipolar probability
measure according to the endpoint ${x_\infty} \in [x_+,x_-]$
\debut \non
\PR_{{x_0}, [x_+,x_-]}^0 \; = \; \int_{x_+}^{x_-} \ud {x_\infty} \,
    A({x_\infty}) \PR^0_{{x_0}; {x_\infty}}
\textrm{ ,}
\fin
where $A$ is the probability density for LERW to end at ${x_\infty}$
\debut \non
A({x_\infty}) & \; = \; & \lim_{\delta, \delta' \downarrow 0}
    \frac{\frac{1}{2 \delta'} \; H({x_0} + i \delta;
        [{x_\infty} - \delta', {x_\infty} + \delta'])}
      {H({x_0} + i \delta); [x_+,x_-])}
    \; = \; \frac{\half Z^{{x_0}; {x_\infty}}}{Z^{{x_0};[x_+,x_-]}_0} \\
& \; = \; & \frac{({x_0} - x_-)({x_0} - x_+)}{(x_- - x_+)({x_\infty} - {x_0})^2}
\textrm{ .}
\fin
As this is just a ratio of the correlation functions, we may say that
the dipolar boundary changing operators are $\psi^+({x_0})$ and
$\half \int_{x_+}^{x_-} \psi^-({x_\infty}) \, \ud {x_\infty}$.
Indeed, the partition function is reproduced by
\debut \non
\vev{\psi^+({x_0}) \, \big( \half \int_{x_+}^{x_-} \psi^-({x_\infty}) \; \ud {x_\infty} \big)}
    \; = \; \frac{1}{\pi} \frac{x_- - x_+}{(x_- - {x_0})(x_+ - {x_0})}
    \; = \; Z^{{x_0}; [x_+,x_-]}_0
\textrm{ .}
\fin

\subsubsection{Field theory representation of Brownian local time}
\label{sec: CFT perturbing operator}
\begin{figure}
\begin{center}
\includegraphics[width=0.9\textwidth]{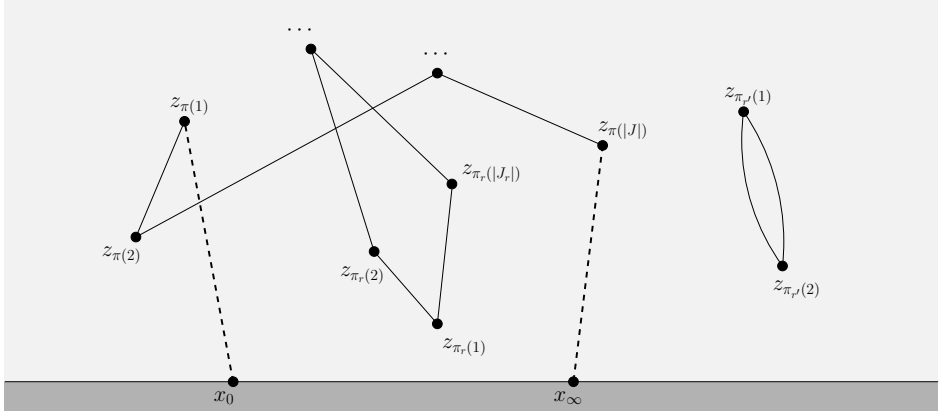}
\caption{\emph{Example diagram for a term appearing in the
correlation function (\ref{eq: chichi correlation function}) of
fields $\ltH$. The component containing boundary points $x_0$ and
$x_\infty$ corresponds to $J$ whereas the rest of the connected
components correspond to $J_1, \ldots, J_s$. Within each component
$J_r$ we sum over inequivalent cyclic orderings $\pi_r$ of it.}}
\label{fig: field theory diagrams}
\end{center}
\end{figure}

In section \ref{sec: local time correlations} we derived the
expressions (\ref{eq: multipoint function dipolar}) and
(\ref{eq: multipoint function chordal}) for Brownian local time
correlations. We recall that in the chordal case, the multipoint
correlation function in the upper half plane is
\debut \non
& & C^{{x_0} ; {x_\infty}} (z_1, \ldots, z_n) \\
\non
& = & \sum_{\pi \in S_n} K_\bH ({x_0}, z_{\pi(1)}) \; \Big( \prod_{j = 2}^n
 G_\bH (z_{\pi(j-1)}, z_{\pi(j)}) \Big) \; K_\bH ({x_\infty}, z_{\pi(n)})
\textrm{ .}
\fin
The two point functions of symplectic fermions involve the same
building blocks
$\vev{\chi^+(z) \chi^-(w)} = G_\bH(z,w)$ and
$\vev{\psi^+(x) \chi^-(z)} = \vev{\chi^+(z) \psi^-(x)} = K_\bH(x;z)$.
Thus the formula is clearly reminiscent of what Wick's formula gives
for correlations of the composite operator 
\debut
\label{eq: local time operator in D}
\ltD(z) = \lim_{z',z'' \rightarrow z} \Big( \chi^-(z') \chi^+(z'')
    - G_\bD (z',z'') \Big)
\textrm{ ,}
\fin
where we substract the full two point function in domain $\bD$
so that in the Wick's
formula no terms with pairing within normal orderings
appear.\footnote{This domain dependent normal ordering
(\ref{eq: local time operator in D}) is not a very natural thing to
do in field theory, but it has the advantage of simplifying the
Wick's formula.}
Inserting also the boundary changing operators
$\psi^+({x_0})$ and $\psi^-({x_\infty})$ for the chordal case, we get
\debut \label{eq: chichi correlation function}
& & \vev{\psi^+({x_0}) \; \ltH(z_1) \; \cdots \; \ltH(z_n) \; \psi^-({x_\infty})} \\
\non
& = & \sum_{\substack{J \subset \{1,\ldots,n\} , \; \pi \in S_J \\
    J_1, \ldots, J_s \textrm{ partition of } \complement J \\
    \textrm{$\pi_r$ cyclic ordering of $J_r$, $r=1,\ldots,s$} }} 
    \prod_{r=1}^s \Big( G_\bH (z_{\pi_r(1)}, z_{\pi_r(2)}) \cdots
    G_\bH (z_{\pi_r(|J_r|)}, z_{\pi_r(1)}) \Big) \\
\non
& & \qquad \times \Big( K_\bH({x_0},z_\pi(1)) \big( \prod_{j = 2}^{|J|}
    G_\bH (z_{\pi(j-1)}, z_{\pi(j)}) \big)
    K_\bH({x_\infty},z_{\pi(|J|)}) \Big)
\textrm{ ,}
\fin
which is represented diagrammatically in figure
\ref{fig: field theory diagrams}.
The terms with $J = \{ 1,2,\ldots,n \}$ are what appear in the
correlation function (\ref{eq: multipoint function chordal}) and
what are illustrated in figure \ref{fig: local time diagram},
the rest of the terms
correspond to disconnected diagrams. To cure this, we must divide
out a loop soup contribution that cancels the disconnected diagrams.
We indeed have
\debut \non
Z^{{x_0}; {x_\infty}}_\nu & = & \lim_{\delta \rightarrow 0} \frac{1}{\delta \, \delta'}
    \EX^{{x_0} + i \delta}_{\BM}
    \Big[ e^{-\eps \int_\bH \tilde{\nu}(z) \ell (z) \; \ud^2 z}
    \; \ind_{{\Bplan}_{\tau_\bH} \in [{x_\infty}-\delta',{x_\infty}+\delta']} \Big] \\
\non
& = & \frac{\vev{\psi^+({x_0}) \; e^{-\eps \int_\bH \tilde{\nu}(z)
	  \ltH(z) \; \ud^2 z} \; \psi^-({x_\infty})}}
    {\vev{e^{-\eps \int_\bH \tilde{\nu}(z) \ltH(z)\; \ud^2 z}}}
\fin
in the sense of formal expansion in powers of $\eps$. In this formula,
however, the precise normal ordering prescription of $\chi^- \chi^+$
doesn't matter: had we made another substraction of the logarithmic
divergence, the result would differ by a constant and would cancel in the
ratio
\debut
Z^{{x_0}; {x_\infty}}_\nu & = &
    \frac{\vev{\psi^+({x_0}) \; e^{-\eps \int_\bH \tilde{\nu}(z)
	  \lt(z) \; \ud^2 z} \; \psi^-({x_\infty})}}
    {\vev{e^{-\eps \int_\bH \tilde{\nu}(z) \lt(z)\; \ud^2 z}}}
    \label{localchi}
\textrm{ ,}
\fin
so in particular we may use the ordinary normal ordering prescription.

From the chordal case formulas (\ref{eq: multipoint function chordal})
and (\ref{localchi}) we can immediately derive also a CFT formula
for the dipolar case by observing that
$\int_{x_+}^{x_-} K_\bH ({x_\infty} , z) \; \ud {x_\infty} = 2 \, H_\bH(z;[x_+,x_-])$.
This reads
\debut \non
Z^{{x_0}; [x_+,x_-]}_\nu & = &
    \frac{ \vev{\psi^+({x_0}) \; e^{-\eps \int_\bH \tilde{\nu}(z)
	  \lt(z) \; \ud^2 z} \; \big( \half \int_{x_+}^{x_-} \psi^-({x_\infty})
          \; \ud {x_\infty} \big) }}
    {\vev{e^{-\eps \int_\bH \tilde{\nu}(z) \lt(z)\; \ud^2 z}}}
\textrm{ .}
\fin

\subsubsection{SLE martingales from conformal field theory}
\label{sec: CFT martingales}
By a two step averaging argument one can construct tautological
martingales for growth processes describing random curves, see
for example \cite{BBH-dipolar_SLE, BBK-multiple_SLEs}.
One splits the full statistical average to average over
configurations that produce a given initial segment of a curve
$\gamma[0,t]$, which is then still to be averaged over all possible
initial segments. The information about the initial segment is
precisely what the SLE filtration $\sF_t$ represents.
If the statistical average can be replaced by CFT correlation
function in the continuum limit, one concludes that for any CFT
field $\sO$ (e.g. product of several primary fields
$\sO = \Phi_{\alpha_1}(z_1,\cc{z}_1) \cdots
\Phi_{\alpha_n}(z_n,\cc{z}_n)$) the ratio
\debut \non
\frac{{\vev{ \sO \; (\bcond) }}_{\bH_t}}{{\vev{ (\bcond) }}_{\bH_t}}
\fin
is a martingale, where $\vev{\cdots ({\bcond})}_{\bH_t}$ represents
CFT expectation in domain $\bH_t = \bH \setminus \gamma[0,t]$ with
insertions of boundary changing operators to account for the boundary
conditions. In the denominator, the expected value of 
the boundary operators corresponds to the partition function.
We emphasize that the operator $\sO$ is constant in time
--- time dependency arises only through the changing domain $\bH_t$ and the
operator placed at the tip $\gamma_t$ of the curve. For example
attempting to use
$\sO = \no{\chi^- \chi^+}_{\bH_t}$ (the closest analog in field theory
of the local time $\ell (z)$) will
not result in a (local) martingale because the normal ordering
(subtraction) is time dependent.

The above argument has a converse, too.
If one considers SLE variant with driving
process $\ud \xi_t = \sqrt{\kappa} \; \ud B_t +
\partial_\xi \log Z_\bH\; \ud t$ and uses transformation
properties of CFT fields, then by a direct check
one concludes that ratios $\vev{\sO}^{\bcond}/Z$ are local
martingales provided the boundary changing operators include a field
$\psi$ at the tip $\gamma_t$ that has a vanishing descendant
$(-4 \, L_{-2} + \kappa \, L_{-1}^2) \psi = 0$.

\bigskip

For the continuum limit of chordal and dipolar LERWs we
have identified the appropriate boundary changing operators in section
\ref{sec: boundary operators} and therefore the ratios
\debut \non
\frac{\vev{\sO \; \psi^+(\gamma_t) \, \psi^-({x_\infty}) }_{\bH_t}}
    {\vev{\psi^+(\gamma_t) \, \psi^-({x_\infty}) }_{\bH_t}}
& \quad \textrm{ and } \quad &
\frac{\vev{\sO \; \psi^+(\gamma_t) \; \int_{x_+}^{x_-}
        \psi^-({x_\infty}) \, \ud {x_\infty} }_{\bH_t}}
    {\vev{\psi^+(\gamma_t) \; \int_{x_+}^{x_-} \psi^-({x_\infty})
        \, \ud {x_\infty}}_{\bH_t}}
\fin
should produce martingales in the two cases respectively.
We will in particular be interested in inserting the perturbing
operator $\lt(z)$, since
to first order the correlation functions in presence of the
perturbation are given by extra insertion of
$\sO = (\unit - \eps \int_\bH \tilde{\nu}(z) \lt(z) \; \ud^2 z)$.

In the half-plane Wick's theorem gives the result
\debut \non
\half \; \vev{ \ltH (z) \; \psi^+(\gamma_t) \;
    \int_{x_+}^{x_-} \psi^-({x_\infty}) \, \ud {x_\infty}}_{\bH} 
& \; = \; & K_{\bH}({x_0};z) \, H_\bH(z;[x_+,x_-])
\fin
which we recognize also as 
the local time correlation function $C^{{x_0}; [x_+,x_-]}(z)$, because
the one point function has no disconnected diagrams.\footnote{We
can then use $\ltH(z) = \lt(z) + \frac{1}{\pi} \log | z - \cc{z} |$
to compute the one point function
$\vev{ \lt (z) \; \psi^+(\gamma_t) \;
\half \int_{x_+}^{x_-} \psi^-({x_\infty}) \, \ud {x_\infty}}_{\bH}
= K_{\bH}({x_0};z) \, H_\bH(z;[x_+,x_-]) - Z^{{x_0}; [x_+,x_-]}_0
\times \frac{1}{\pi} \log | z - \cc{z} |$.}
Recall the logarithmic anomaly in the transformation property of
$\lt(z)$, eq.(\ref{eq:loganom}) to get 
\debut \non
& & N_t^{{x_0};[x_+,x_-]}(z) \; = \; \frac{\vev{\lt(z) \; \psi^+(\gamma_t) \;
        \int_{x_+}^{x_-} \psi^-({x_\infty}) \, \ud {x_\infty} }_{\bH_t}}
    {\vev{\psi^+(\gamma_t) \; \int_{x_+}^{x_-} \psi^-({x_\infty})
        \, \ud {x_\infty}}_{\bH_t}} \\
\non
& \; = \; & \frac{\vev{\Big( \lt (g_t(z)) + \frac{1}{\pi} \log |g'_t(z)| \Big)
    \; \psi^+(\xi_t) \;
    \int_{X^+_t}^{X^-_t} \psi^-({x_\infty}) \, \ud {x_\infty} }_{\bH}}
    {\vev{\psi^+(\xi_t) \; \int_{X^+_t}^{X^-_t} \psi^-({x_\infty})
        \, \ud {x_\infty}}_{\bH}} \\
\label{eq: one point function martingale}
& \; = \; & -\frac{1}{\pi} \log \frac{2 \; \im ( g_t(z) )}{|g_t'(z)|}
    + \frac{K_{\bH} (\xi_t ; g_t(z)) \, H_{\bH}(g_t(z) ; [X^+_t,X^-_t])}
        {Z_0^{\xi_t ; [X^+_t,X^-_t]}}
\textrm{ ,}
\fin
where $X^\pm_t = g_t(x_\pm)$. The process $N_t(z)$ should be a local
martingale by
construction and one can indeed verify this directly by It\^o's
formula.

The formula (\ref{eq: one point function martingale}) has a natural
probabilistic interpretation, too: as a conditional expected value
of the local time of the underlying random walk at $z$.
The two parts correspond to the splitting
$\ell_\bH = \ell_{\bH \setminus \gamma[0,t]} + \ell^{(t)}_\bH$.
The second term is indeed,
by conformal invariance of the local time $\ell(z)$, just the
expected value of the local time of Brownian motion in
$\bH \setminus \gamma[0,t]$ started from $\gamma_t$ and conditioned
to exit through $[x_+,x_-]$.
The first term is $A_t = -\frac{1}{\pi} \log \rho_{\bH_t}(z)$, where
$\rho_{\bH_t}(z) = \frac{2 \, \im(g_t(z))}{|g_t'(z)|}$ is the conformal
radius of $z$ in $\bH \setminus \gamma[0,t]$. In particular the
first term is an increasing process.
Recall that in the discrete setup, conditional on loop erasure
producing a given initial segment
of the curve, the second term corresponds to the expected local time at
$z$ of the underlying random walk after its last visit to the tip of the
curve, see \cite{LSW-LERW_and_UST} whereas the first part,
more precisely $A_t-A_0$,
should be interpreted as the expected
local time at $z$ of the (erased) loops until the last visit to the tip.
The fact that
$A_t$ is increasing is then natural since as time increases we
erase more loops.
Seen this way, $A_t - A_0$ is also what we called the (nonlocal)
interface energy of the LERW in section \ref{sec: interface energy}.

It has been argued \cite{LSW-conformal_restriction}
that it should be possible  to add to SLE${}_2$ Brownian bubbles so as to
reconstruct the underlying Brownian motion. We notice indeed that
\debut \non
A_t - A_0 = \frac{4}{\pi} \; \int_0^t \frac{\big( \im g_s(z) \big)^2}
        {|g_s(z)-\xi_s|^4} \; \ud s
\textrm{ ,}
\fin
where the integrand is morally twice the ``expected'' local time at
$z$ of a Browian bubble in $\bH \setminus \gamma[0,s]$ from $\gamma_s$.
Actually Brownian bubbles don't form a probability measure but an
infinite measure. If we normalize it as in
\cite{LSW-conformal_restriction} (but we must not forget about the time
parametrization of the bubbles, see \cite{LW-Brownian_loop_soup}),
the integral with respect to the bubble
measure of the local time is 
$\frac{\pi}{2} K(\xi_s ; g_s(z))^2 =
\frac{2}{\pi} \big( \im g_s(z) / |g_s(z)-\xi_s|^2 \big)^2$.
The factor two is an intensity at which we need to add the bubbles
to the curve --- it is minus the central charge, $\lambda = -c = 2$.

\bigskip

In the chordal case we obtain similar formulas --- in fact they can
also be recovered by limit of the dipolar case. For the record, we
give the (local) martingale
\debut \non
N_t^{{x_0} ; {x_\infty}} (z)
& \; = \; & -\frac{1}{\pi} \log \frac{2 \; \im ( g_t(z) )}{|g_t'(z)|}
    + \frac{K_{\bH} (\xi_t ; g_t(z)) \, K_{\bH}(\eta_t; g_t(z))}
        {Z_0^{\xi_t ; \eta_t}}
\textrm{ ,}
\fin
where $\eta_t = g_t({x_\infty})$. It is in particular worth noticing that
the ``expected local time of the erased loops'' $A_t-A_0$ has the
same formula and depends only on the shape of the ``initial segment
of the loop-erasure'' $\gamma[0,t]$.

\subsection{Off-critical LERW and massive symplectic fermions}
\label{sec: offcritical field theory}

The conformal field theory of LERW is the symplectic fermion
theory with central charge $c=-2$. As we have argued when defining 
the scaling limit of the LERW, going off-criticality amounts to perturbing
by an operator of scaling dimension $0$. In terms of Brownian motion
the off-critical weighting is given by the local time which is closely linked
to the composite operator $:\chi^-\chi^+:$ as we've shown above,
cf. eq.(\ref{localchi}) and $N_t(z)$ in section \ref{sec: CFT martingales}.
In fact, as the perturbing field is $:\chi^+\chi^-:$ and the action
for the off-critical theory thus reads
$$(const.) \int_\bH d^2z [J_{\alpha\beta}\bar\partial\chi^\alpha \partial\chi^\beta  
+ 8\nu(z)\,J_{\alpha\beta}\chi^\alpha \chi^\beta ],$$
the need to divide by $\vev{e^{-\int \nu(z) \lt(z) \; \ud^2 z}}$
stems just from the normalization of the new measure.
We remark in particular that the off-critical theory is still Gaussian
with two point function
\debut \non
\vev{\chi^\alpha (z,\cc{z}) \chi^\beta (w,\cc{w})}_\nu
    = J^{\alpha \beta} G^\nu_\bH(z,w)
\textrm{ ,}
\fin
where $(- \nu(w) + \half \lapl_w) G^\nu_\bH(z,w)=-2\delta(z-w)$.

For simplicity we look at the theory
in the upper half plane. The boundary conditions are identical to
that of the critical theory.

\bigskip

Suppose, as has been argued, that the off-critical measure
$\PR^\nu_{\bH}$ on curves differs from the critical one by a
Radon-Nikodym derivative $M_t$ given by (\ref{Mcontinuous}).
We have been able to compute the limit of partition functions in
(\ref{eq: dipolar perturbative expansion}) \&
(\ref{eq: multipoint function dipolar}) or alternatively in
(\ref{localchi}), and we know that the energy term
$\Delta {\cal E}_\bH(\gamma[0,t])$ is monotone in $t$, thus of finite
variation (can not have a $\ud B_t$ like increment). This is
in fact enough to determine what the energy term is in our case:
it must compensate the drift so that $M_t$ becomes a martingale
and it is not difficult to check that this requires
\debut \non
e^{\Delta {\cal E}_\bH(\gamma[0,t])} \; = \; &
\frac{\bvev{\exp \big( -\int \ud^2 z \; \nu(z) \ltH (z) \big) \;
    (\bcond)}_{\bH_t}}
    {\bvev{\exp \big( -\int \ud^2 z \; \nu(z) \lt_{\bH_t} (z) \big) \;
    (\bcond)}_{\bH_t}} \\
= \; & \exp \Big( - \frac{1}{\pi} \int  \ud^2 z \; \nu(z) \;
    \log \big( \frac{\rho_{\bH_t}(z)}{\rho_\bH(z)} \big) \Big)
\textrm{ .}
\fin
The change in interface energy is therefore given by the bubble
soup $A_t-A_0$ as we could have expected.
Furthermore and importantly, the field theoretic formula
(\ref{eq: M from QFT}) for
$M_t$ to first order holds with $\Phi(z) = \lt(z)$.

\subsubsection{Subinterval hitting probability from field theory}
\label{sec: CFT subinterval hitting}

We will now show how to use the field theory interpretation to compute
probabilities for the off-critical LERW. We work in the dipolar setup,
a LERW from ${x_0}$ to $[x_+,x_-]$ in $\bH$,
and ask what is the probability for the
endpoint of the LERW to be on a subinterval $S=[x'_+,x'_-] \subset [x_+,x_-]$.
In the next section we derive the same result from direct probabilistic
considerations.

From Boltzmann rules, this probability is the ratio of two partition functions:
the partition of LERW exiting on $[x'_+,x'_-]$
by that of LERW exiting on $[x_+,x_-]$.
In field theory this becomes the ratio of two correlation functions but
with different boundary conditions (or equivalently, insertion of boundary 
changing operators at different locations). Hence, this hitting probability
is expected to be:
$$
\PR^\nu_{\xi_0 ; [x_+,x_-]} \Big[ \textrm{end in $[x'_+,x'_-]$} \Big]=
\frac{\vev{\psi^+(\xi_0)\, \half \big( \int_{x'_+}^{x'_-} \psi^-({x_\infty}) \ud {x_\infty} \big) }_\nu}
{\vev{\psi^+(\xi_0)\, \half \big( \int_{x_+}^{x_-} \psi^-({x_\infty}) \ud {x_\infty} \big) }_\nu}
$$
where the operator $\psi^+(\xi_0)$ is the operator which creates the LERWs
and the operator $\half \int_{x'_+}^{x'_-} \psi^-(x_\infty) \, \ud x_\infty$
or $\half \int_{x_+}^{x_-} \psi^-(x_\infty) \, \ud x_\infty$
are those conditioning
the curves to stop on the interval $[x'_+,x'_-]$ or $[x_+,x_-]$, so that they
impose the boundary conditions.

At criticality, the correlation function
$\vev{\psi^+(\xi_0) \, \half \int_{x'_+}^{x'_-}
\psi^-(x_\infty) \, \ud x_\infty}_0$
is computable from the limit behavior of the harmonic measure 
$H_\bH(z_0,[x'_+,x'_-])$ as $z_0\to \xi_0$ so that
$$\vev{\psi^+(\xi_0)\, \big( \int_{x'_+}^{x'_-} \psi^-({x_\infty}) \ud
  {x_\infty} \big)}_0= \frac{2}{\pi} \frac{(x'_- -
  x'_+)}{(\xi_0-x'_-)(\xi_0-x'_+)} \textrm{ .}$$ 
The hitting probability in $\bH$ at the conformal point is thus
$$\PR^0_{\xi_0 ; [x_+,x_-]} \Big[ \textrm{end in $[x'_+,x'_-]$} \Big]=
\frac{(x'_- - x'_+)(\xi_0-x_-)(\xi_0-x_+)}{(x_- - x_+)(\xi_0-x'_-)(\xi_0-x'_+)} .$$

Off-criticality, the correlation functions
$\vev{\psi^+(\xi_0)\, \half \big( \int_{x'_+}^{x'_-} \psi^-({x_\infty}) \ud {x_\infty} \big)}_\nu$
are computable via the limiting behavior of 
$\vev{\chi^+(z_0)\, \half \big( \int_{x'_+}^{x'_-} \psi^-({x_\infty}) \ud {x_\infty} \big)}_\nu$
with $z \in \bH$ but approaching the real axis, $z \to \xi_0$.
The off-critical probability is thus expected to be
\begin{eqnarray}
\PR^\nu_{\xi_0 ; [x_+,x_-]} \Big[ \textrm{end in $[x'_+,x'_-]$} \Big]=
 \lim_{z\to \xi_0}\frac{\Gamma^\nu_{\bH, [x'_+,x'_-]}(z)}{\Gamma^\nu_{\bH, [x_+,x_-]}(z)}
 \label{field_proba}
 \end{eqnarray}
with $\Gamma^\nu_{\bH, [x'_+,x'_-]}(z)$ solution of 
$(- \nu(z) + \half \lapl_z) \Gamma^\nu_{\bH, [x'_+,x'_-]}(z)=0$.
To find the boundary conditions observe that the leading term in
the OPE
$\chi^+(z) \psi^-({x_\infty}) \sim \frac{-2}{\pi}
\im \big( \frac{1}{z-{x_\infty}}\big)$ remains unchanged in the
off-critical theory. One finds that
$\Gamma^\nu_{\bH, [x'_+,x'_-]}(z)\to1$ for $z \to (x'_+,x'_-)$
and $\Gamma^\nu_{\bH, [x'_+,x'_-]}(z)\to0$ for $z\to \bR \setminus [x'_+,x'_-]$.
Both $\Gamma^\nu_{\bH, [x'_+,x'_-]}$ and $\Gamma^\nu_{\bH, [x_+,x_-]}$
vanish at $\xi_0 \in \bR \setminus [x_+,x_-]$,
but their ratio tends to a finite limit.

In the following section, we shall present a probabilistic derivation
of this  field theory inspired formula.

\subsubsection{Probabilistic derivation of subinterval hitting probability}
Above we gave a field theory flavoured discussion of the probability
that a perturbed LERW in $\bH$ from ${x_0}$ to $[x_+,x_-]$ ends on
a subinterval $[x'_+,x'_-] \subset [x_+,x_-]$. It is easy to justify the
formulas obtained there
by computations with Brownian motion.

Most importantly, we notice that the question of endpoint is a property
of the (weighted) random walk $W$ that we then decided to loop erase.
Indeed, by construction the loop erasing procedure doesn't change the
starting point and end point. Therefore we only need to find the
subinterval hitting probability of the weighted random walk, which in
the continuum boils down to a Brownian motion computation. 

We thus consider a walk in the upper half plane,
started from
$x_0 \in \bR$ (or an approximation to it) and conditioned to
exit the half plane through $[x_+,x_-] \subset \bR$ (a lattice
approximation of it). The walk
$(W^{(a)}_j)_{j=0}^{\tau_{\bH^{(a)}}}$ is
weighted by $\exp ( - \sum_j a^2 \nu^{(a)}(W^{(a)}_j) )$ relative to
the symmetric random walk, so the probability of an event $E$ is
\debut \non
\PR^{\nu}_{w;[x_+,x_-]} [ E ] \; = \;
    \frac{\EX^w_{\RW} \big[ \ind_{E \, \cap \,
        \{ W^{(a)}_{\tau_{\bH^{(a)}}} \in [x_+,x^-] \}} \;
        \exp ( - \sum_j a^2 \nu^{(a)}(W^{(a)}_j) ) \big]}
    {\EX^w_{\RW} \big[ \ind_{W^{(a)}_{\tau_{\bH^{(a)}}} \in [x_+,x^-]} \;
        \exp ( - \sum_j a^2 \nu^{(a)}(W^{(a)}_j) ) \big]}
\textrm{ .}
\fin
Take $E$ to be the event $W^{(a)}_{\tau_{\bH^{(a)}}} \in [x'_+,x'_-]
\subset [x_+x_-]$. The continuum limit $a \downarrow 0$ of the
probability of exiting through $[x'_+,x'_-] \subset [x_+,x_-]$ is then
computed using Brownian motion
\debut \label{eq: weighted conditional hitting}
\PR^\nu_{w ; [x_+,x_-]} \big[ \textrm{end in $[x'_+,x'_-]$} \big] & = &
\frac{\EX_{\BM}^{w} [ e^{-\int \nu({\Bplan}_s) \ud s} \;
    \ind_{{\Bplan}_{\tau_\bH} \in [x'_+,x'_-]} ] }
    {\EX_{\BM}^{w} [ e^{-\int \nu({\Bplan}_s) \ud s} \;
    \ind_{{\Bplan}_{\tau_\bH} \in [x_+,x_-]} ] }
\textrm{ .}
\fin
If $\nu \equiv 0$, the numerator and denominator are nothing but
the harmonic measures of $[x'_+,x'_-]$ and $[x_+,x_-]$ respectively, and
the field theoretic formula at criticality is justified.
For non-zero $\nu$,
we can get a partial differential equation for the numerator
and denominator by Feynman-Kac formula: denoting
$\Gamma(w) = \EX_{\BM}^{w} [ e^{-\int \nu({\Bplan}_s) \ud s} \;
\ind_{{\Bplan}_{\tau_\bH} \in [x'_+,x'_-]} ]$ we get a martingale
\debut
\non
& & \EX_{\BM}^{w} \big[ \exp ( -\int_0^{\tau_\bH} \nu({\Bplan}_s) \ud s ) \;
    \ind_{{\Bplan}_{\tau_\bH} \in [x'_+,x'_-]}
    \; \big| \; \sF^{\BM}_{t \minim \tau_\bH} \big] \\
\non
& = \; & \exp ( -\int_0^{t \minim \tau_\bH} \nu({\Bplan}_s) \ud s ) \; \times \;
    \tilde{\EX}_{\BM}^{{\Bplan}_{t \minim \tau_\bH}}
    \big[ \exp ( -\int_0^{\tilde{\tau}_\bH} \nu(\tilde{\Bplan}_s) \ud s ) \;
    \ind_{\tilde{\Bplan}_{\tilde{\tau}_\bH} \in [x'_+,x'_-]} \big] \\
\non
& = \; & \exp ( -\int_0^{t \minim \tau_\bH} \nu({\Bplan}_s) \ud s ) \; \times \;
    \Gamma({\Bplan}_{t \minim \tau_\bH})
\fin
and the requirement for the It\^o drift of this to vanish is
\debut \non
0 & \; = \; & - \nu(w) \Gamma(w) + \half \lapl_w \Gamma(w)
\textrm{ .}
\fin
This is supplemented by the boundary conditions that are obvious from
the definition of $\Gamma$
\debut \non
\Gamma(w) \rightarrow \left\{ \begin{array}{ll}
0 \qquad & \textrm{ as $w \rightarrow \bR \setminus [x'_+,x'_-]$} \\
1 \qquad & \textrm{ as $w \rightarrow (x'_+,x'_-)$}
\end{array} \right.
\textrm{ .}
\fin
The ratio (\ref{eq: weighted conditional hitting}) is thus just what
we argued from field theory.

\bigskip

If we are interested in small perturbations, it is useful
to take $\nu = \eps \tilde{\nu}$ and write the solution
$\Gamma$ as a power series in $\eps$
\debut \non
\Gamma (w) = \sum_{k=0}^\infty \eps^k \Gamma_k (w)
\textrm{ .}
\fin
The zeroth and first orders are explicitly
\debut
\non
\Gamma_0 (w) & = & \frac{1}{\pi} \im \log \frac{w-b}{w-a}
    \, = \, H_{\bH}(w ; [a,b]) \\
\non
\Gamma_1 (w) & = & - \int_\bH \ud^2 z \;
    \tilde{\nu}(z) H_{\bH}(z;[a,b]) G_\bH(w,z)
\textrm{ .}
\fin
Furthermore we want the walk to start from the boundary, at
$x_0$. The limit $w \rightarrow x_0$ of $\Gamma$ vanishes, but
the ratio (\ref{eq: weighted conditional hitting}) remains
finite. We find that the probability to end in
$[x'_+,x'_-] \subset [x_+,x_-]$ is, to first order in $\eps$,
given by
\debut \label{eq: subinterval hitting first order}
& & \PR^\nu_{x_0 ; [x_+,x_-]} \Big[ \textrm{end in $[x'_+,x'_-]$} \Big] \\
\non
& \approx \; & \frac{Z_0^{{x_0};[x'_+,x'_-]}}{Z_0^{{x_0};[x_+,x_-]}}
    \; +\; \eps \, \int_\bH \ud^2 z \; \tilde{\nu}(z) \, K_{\bH}({x_0};z) \,
    \Big\{ - \frac{H_\bH (z;[x'_+,x'_-])}{Z^{{x_0};[x_+,x_-]}} \\
\non
& & \qquad + \frac{Z^{{x_0}; [x'_+,x'_-]} \; H_\bH (z; [x_+,x_-])}
    { ( Z^{{x_0};[x_+,x_-]} )^2} \Big\} \\
\non
& \approx \; & \frac{(x'_+,x'_-) \; (x_+ - x_0) (x_- - x_0)}
    {(x_- - x_+) \; (x'_+ - x_0)(x'_- - x_0)} \\
\non
& & + \eps \; \frac{2}{\pi} \;
    \int_\bH \ud^2 z \; \tilde{\nu}(z) \; \im \big(
    \frac{1}{z-x_0} \big) \; \Big\{
    \frac{(x_+ - x_0) (x_- - x_0)}{x_- - x_+}
        \; \im \big( \log \frac{z-x'_+}{z-x'_-} \big) \\
\non
& & \qquad - \frac{(x'_+ - x'_-) \; (x_+ - x_0)^2 (x_- - x_0)^2}
    {(x_- - x_+)^2 \; (x'_+ - x_0)(x'_- - x_0)}
        \; \im \big( \log \frac{z-x_-}{z-x_+} \big) \Big\}
\textrm{ .}
\fin

\subsection{Link with perturbed SLEs}
\label{sec: perturbed SLEs}

\subsubsection{Perturbation to driving process using hitting distribution}
\label{sec: subinterval hitting}

Suppose that the perturbed LERW has a continuum limit that is
absolutely continuous with respect to SLE${}_2$ (for us what is
important is that the measures on initial segments $\gamma[0,t]$
are absolutely
continuous so that the driving processes differ only by a drift term).
We can then describe the
curve in the continuum limit by a Loewner chain
$(g_t)_{t \in [0,T]}$,
whose driving process would solve a stochastic differential equation
\debut
\ud \xi_t \; = \; \sqrt{\kappa} \, \ud B_t + F_t \, \ud t 
\textrm{ .}
\fin
The drift $F_t$ would depend on $\gamma[0,t]$ and
$\nu |_{\bH \setminus \gamma[0,t]}$.

We can use the event $\gamma_T \in [x'_+,x'_-]$ to build the
martingale
\debut \non
P_t \; = \; \EX^\nu_{x_0;[x_+,x_-]} \big[ \ind_{\gamma_T \in [x'_+,x'_-]} \; \big| \;
    \sF^\gamma_t \big]
\fin
so that
\debut \non
P_0 = \PR^{\nu}_{x_0 ; [x_+,x_-]} [\textrm{hit $[x'_+,x'_-]$}]
\textrm{ .}
\fin
The $P_t$ is the conditional probability, given $\gamma[0,t]$,
to hit $[x'_+,x'_-]$. By the Markov property of the perturbed LERW,
this is the probability to hit $[x'_+,x'_-]$ for a LERW in
$\bH \setminus \gamma[0,t]$ from $\gamma_t$ to $[x_+,x_-]$,
perturbed with $\nu |_{\bH \setminus \gamma[0,t]}$.
Conformal invariance of the underlying Brownian motion
allows us to write this as
\debut \non
P_t = \PR^{\nu_t}_{\xi_t ; [X^+_t, X^-_t]}
    [\textrm{hit $[g_t(x'_+),g_t(x'_-)]$}]
\textrm{ ,}
\fin
where $\nu_t (z) = |(g_t^{-1})'(z)|^2 \, \nu(g_t^{-1}(z))$
because of the appropriate time change (see section
\ref{sec: conformal maps of local time})
and $X^\pm_t = g_t(x_\pm)$.
From the formula (\ref{eq: subinterval hitting first order})
of the previous section we find, to first order in $\eps$,
\debut \non
P_t & \approx \; & \frac{Z^{\xi_t;[g_t(x'_+),g_t(x'_-)]}}{Z^{\xi_t;[X^+_t,X^-_t]}}
    \; +\; \eps \, \int_{\bH \setminus \gamma[0,t]} \ud^2 z \;
    \tilde{\nu}(z) \, K_{\bH}(\xi_t;g_t(z)) \, \\
\non
& & \quad \times \; \Big\{
    \frac{Z^{\xi_t; [g_t(x'_+),g_t(x'_-)]} \; H_\bH (g_t(z); [X^+_t,X^-_t])}
    { ( Z^{\xi_t;[X^+_t,X^-_t]} )^2}
    - \frac{H_\bH (g_t(z);[g_t(x'_+),g_t(x'_-)])}{Z^{\xi_t;[X^+_t,X^-_t]}} \Big\}
\textrm{ .}
\fin
Since this should be a martingale, we require its It\^o drift to
vanish. To zeroth order in $\eps$ we get just
$F_t \approx \frac{-2}{\xi_t - X^+_t} + \frac{-2}{\xi_t - X^-_t}
    + \Order(\eps)$,
which says that our curve is an SLE${}_2(-2,-2)$ i.e. a dipolar
SLE${}_2$. A naive computation neglecting the change of domain
of integration and exchanging integral and It\^o differential
shows the effect of the
perturbation at first order in $\eps$
\debut
\non
F_t & \approx & \frac{-2}{\xi_t - X^+_t} + \frac{-2}{\xi_t - X^-_t} 
    \; + \; 4 \eps \, \int_{\bH \setminus \gamma[0,t]} \ud^2 z
    \; \tilde{\nu}(z) \; \frac{H_\bH(g_t(z);[X^+_t,X^-_t])}{X^-_t - X^+_t} \\
& & \qquad \; \times \;
        \im \big( \frac{(g_t(z)-X^-_t)(g_t(z)-X^+_t)}
        {(g_t(z)-\xi_t)^2} \big) 
\textrm{ .}
\fin

\bigskip

We have thus found out what is the first order correction to driving
process by using the subinterval hitting probabilities computed in section
\ref{sec: subinterval hitting}. This argument works for LERW aimed
towards a nondegenerate interval $[x_+,x_-]$, i.e. the dipolar
setting. Chordal case could be obtained from this as a limit,
but it is very instructive to give another argument that can be applied
directly also in the chordal setting and that follows the general strategy
outlined in section \ref{sec: probability measures}.
We will do that next.

\subsubsection{Perturbation to driving process from Girsanov's formula}

We have argued in sections
\ref{sec: probability measures} and \ref{sec: offcritical field theory}
that the continuum offcritical LERW measure should be
absolutely continuous with respect to SLE${}_2$, with Radon-Nikodym
derivative\footnote{The intuition from weighted random walks says
the Radon-Nikodym derivative should be
$\sZ_\nu^{-1} \times \EX \big[ \exp \big( - \int_\bH \ud^2 z \;
\nu(z) \ell(z) \big) \;  \big| \; \sF^\gamma_t \big]$. Using this formula
one arrives at the same conclusion, but from a rigorous point of view
a coupling of the 2-d Brownian motion with ``its loop erasure''
SLE${}_2$ is missing anyway: it is not known how to construct the two
in the same probability space so that the SLE filtration $\sF_t$
and Brownian local time would both make sense.}
(\ref{Mcontinuous})
\debut \non
\frac{\ud \PR^\nu}{\ud \PR^0} \Big|_{\sF_t} & = & M_t \; = \;
    e^{\Delta {\cal E}_\bH(\gamma[0,t])} \;
    \frac{Z_\nu^{\bH\setminus\gamma_{[0,t]} ; \bcond}
    / Z_0^{\bH\setminus\gamma_{[0,t]} ; \bcond} }{
    Z_\nu^{\bH ; \bcond}/Z_0^{\bH ; \bcond} } \\
\non
& = & \const \times \frac{\vev{e^{-\int_{\bH_t} \nu(z) \lt (z) \, \ud^2 z} \;
        (\bcond)}_{\bH_t}}
    {\vev{(\bcond)}_{\bH_t}}
\textrm{ ,}
\fin
where the constant is there just to make the initial value unity,
$M_0=1$. To first order in $\eps$ we have
\debut \non
\frac{\ud \PR^{\eps \tilde{\nu}}}{\ud \PR^0} \Big|_{\sF_t}
& \approx & \frac{1 - \eps \int \tilde{\nu}(z) N_t(z) \; \ud^2 z}
    {1-\eps \int \tilde{\nu}(z) N_0(z) \; \ud^2 z}
\textrm{ ,}
\fin
where $N_t(z)$ is the one-point function martingale of section
\ref{sec: CFT martingales}.
Explicitly in the chordal case we have
\debut \non
& & 1 - \eps \int \tilde{\nu}(z) N^{x_0;x_\infty}_t(z)\; \ud^2 z \\
\non
& \approx & 1 - \eps \; \frac{2}{\pi} \int \ud^2 z \; \tilde{\nu}(z)
    \Big\{ (\eta_t - \xi_t)^2 \; \im \big(\frac{1}{g_t(z)-\xi_t}
    \big) \; \im \big(\frac{1}{g_t(z)-\eta_t} \big) \\
\non
& & \qquad \qquad
    - \frac{1}{2} \log \big( \rho_t(z) \big) \Big\}
\textrm{ .}
\fin

Since $B_t$ appearing in the (critical) chordal driving process
$\ud \xi_t = \sqrt{2} \; \ud B_t + \frac{\rho_c}{\xi_t - \eta_t}
\; \ud t$
is a $\PR^0_{x_0 ; x_\infty}$-Brownian motion, an application of
Girsanov's formula
tells us that under $\PR^{\eps \tilde{\nu}}_{x_0 ; x_\infty}$ it has
additional drift
$$\ud \qv{B \, , \, - \eps \int \ud^2 z \; \tilde{\nu}(z) N(z)}_t$$
(we will exchange integrations and quadratic variations etc. in
good faith).
This means that the driving process $\xi_t$ satisfies
\debut \non
\ud \xi_t & \approx & \sqrt{2} \; \ud B_t'
    + \frac{\rho_c}{\xi_t - \eta_t} \; \ud t \\
\non
& & + 2 \eps \; \int \ud^2 z \; \tilde{\nu}(z) \;
        K_\bH(\eta_t ; g_t(z)) \;
    \im \big( \frac{(g_t(z)-\eta_t)^2}{(g_t(z)-\xi_t)^2} \big)
\fin
with $B_t'$ a
$\PR^{\eps \tilde{\nu}}_{x_0;x_\infty}$-Brownian motion.

\bigskip

In the dipolar setup, we have similarly
\debut \non
& & 1 - \eps \int \tilde{\nu}(z) N^{x_0; [x_+,x_-]}_t(z)\; \ud^2 z \\
\non
& \approx & 1 - \eps \; \frac{2}{\pi} \int \ud^2 z \; \tilde{\nu}(z)
    \Big\{ \frac{(X^-_t - \xi_t)(X^+_t - \xi_t)}{X^+_t - X^-_t} \;
    \im \big(\frac{1}{g_t(z)-\xi_t} \big) \;
    \im \big( \log \frac{g_t(z)-X^-_t}{g_t(z)-X^+_t} \big) \\
\non
& & \qquad \qquad
    - \frac{1}{2} \log \big( \rho_t(z) \big) \Big\}
\textrm{ .}
\fin
As above, in the dipolar driving process
$\ud \xi_t = \sqrt{2} \; \ud B_t + (\frac{\rho_d}{\xi_t - X^-_t} +
\frac{\rho_d}{\xi_t - X^+_t}) \; \ud t$ we have a
$\PR^0_{x_0;[x_+,x_-]}$-Brownian
motion $B_t$. Applying Girsanov's formula again gives us
to first order in $\eps$
\debut \non
\ud \xi_t & \approx & \sqrt{2} \; \ud B_t'
    + \frac{\rho_d}{\xi_t - X^+_t} \; \ud t
    + \frac{\rho_d}{\xi_t - X^-_t} \; \ud t \\
\non
& & + 4 \eps \; \int \ud^2 z \; \tilde{\nu}(z) \;
        \frac{H_\bH(g_t(z);[X^+_t,X^-_t])}{X^-_t - X^+_t} \;
    \im \big( \frac{(g_t(z)-X^-_t) (g_t(z)-X^+_t)}{(g_t(z)-\xi_t)^2}
      \big)
\fin
where $B_t'$ a $\PR^{\eps \tilde{\nu}}$-Brownian motion. In the limit
$X^+_t,X^-_t \rightarrow \eta_t$ we of course recover the chordal
result. The formula also coincides with the offcritical dipolar drift
we got by the subinterval hitting probability argument.

\section{Conclusions}
We have studied the example of off-critical loop-erased random
walk in some detail, discussing the statistical physics, field
theory and probability measure on curves. We have done this in
such a way that it should be easy to see which parts can be
expected to generalize to more physically relevant near-critical
interfaces.

The most important observation is that one may try to use SLE-like
methods to understand interfaces even if the model is not
precisely at its critical (conformally invariant) point. We have
proposed a field theoretical formula for the Radon-Nikodym
derivative of the off-critical measure with respect to the
critical one, which can then by Girsanov's theorem be translated to
a stochastic differential equation for the Loewner driving process.
For off-critical LERW we've given two derivations
of the equation for the driving process and they coincide with the
field theoretic
prediction once the perturbing operator has been identified.
We remark that the off-critical driving process is not Markovian,
it's increments depend in a very complicated manner on its past.
But this must be so, because Loewner's technique reduces the future
of the curve to the original setup by conformal maps and the
off-critical model is not conformally invariant.

We hope that this example encourages studies of interfaces in other
off-critical models, some maybe physically more relevant.
Furthermore, even after the novel connections of LERW to field theory,
there clearly remains important questions to be
understood before we have a fully satisfactory field
theory description of LERWs.

\appendix

\section{Random walks as an example}
\label{app: Girsanov}

The example of random walks in 1D can serve as a trivial illustration
of the themes discussed in this article. Suppose we weight walks on
$\mathbb Z$ starting from $0$ by giving weight $\mu e^\gamma >0$ to
each positive step and $\mu e^{-\gamma} >0$ to each negative step. The
weight of a walk of $n$ steps ending at $s$ is simply $\mu ^n e ^
{s\gamma}$ if $-n \leq s \leq n$ and $n-s$ is even, but $0$
otherwise. The partition function $Z$, obtained by summing over all
paths, converges if and only if $w\equiv 2\mu \cosh \gamma <1$, and
then $Z=\frac{1}{1-w}$. We infer that the average length of a path is
$\mu\frac{\partial \log Z}{\partial \mu}=\frac{w}{(1-w)^2}$ which goes
to $+\infty$ as $w$ approches $1^-$. Hence a critical theory is
obtained for $w=1$ i.e. when the weight of walks of length $n$ is $1$
for each $n$ and the model has a purely probabilistic random walk
description. Hence the critical line is $2\mu \cosh \gamma =1$.

The quadratic fluctuation of the end point of the walk is
$\left(\lambda\frac{\partial }{\partial \lambda}\right)^2 \log Z
=\frac{w-w^2+v^2}{(1-w)^2}$ where $v\equiv 2\mu \sinh \gamma$. For
fixed $v\neq 0$, this blows up like the average length of the walk for
$w\rightarrow 1^-$, but for $v=0$ the divergence is milder. Hence the
point $\gamma=0$, which is nothing but the simple symmetric random
walk, is special among the critical points. At $v=0$, the weight of a
path of length $n$ is simply $(w/2)^n$ and a continuum limits exist
for which $\log (w/2)$ scales like the square of the lattice spacing,
leading in the continuum to weight paths by the local time, as used at
length in these notes in the 2D situation.

But for now, let us concentrate on the critical line. The weight of an
$n$ steps path ending at $S_n$ is $e^{\gamma
S_n}(2\cosh \gamma)^{-n}$ and the ratio of this weight to that of the
simple symmetric random walk is $Q_n=e^{\gamma S_n}(\cosh
\gamma)^{-n}$ which is readily checked to be a martingale for the
simple symmetric random walk. As in the continuum, this martingale can
be used to change the measure to a new one under which the symmetric
random walk is turned to an asymmetric one. We get this trivially in
the discrete setting, but in more complicated situations, the
flexibility offered by the continuum theory and Girsanov's theorem is
invaluable. So we turn to the continuum limit.

Introduce a lattice spacing $a$ so that the macroscopic position after
$n$ steps is $aS_n$. If this has a continuum limit and the martingale
$Q_n$ as well, one must take $\gamma \sim a g$ and, in order for the
second factor $(\cosh \gamma)^{-n}$ to converge, one has to set
$a^2n=t$ and keep $t$ fixed when taking the lattice spacing to $0$,
defining $aS_n\rightarrow X_t$ in the limit. Then $Q_n$ goes to
$M_t=e^{g X_t-\frac{g^2 t}{2}}$. If $X_t$ is a 1D Brownian motion
$dX_t=dB_t$, with quadratic variation $(dB_t)^2=dt$, then $M_t$ is well
known to be a martingale, well-defined for $t<\infty$ finite,
normalized to $M_0=1$ and such that $M_t^{-1}dM_t = g dB_t$.

With respect to the dressed expectation $\hat  \EX[\cdots ]=\EX[\cdots M_t]$, 
the process $X_t$ satisfies $dX_t=d\hat B_t +g dt$ with $\hat B_t$ a Brownian
motion with respect to $\hat  \EX[\cdots ]$.

In particular, it is easy to check that $\EX[X_t\, M_t]=g t$. 
More generally, for any function $F(X_t)$ we have 
$\frac{d}{dt}\hat \EX[F(X_t)]=\hat \EX[{\cal A}\cdot F(X_t)]$ 
with  dressed stochastic evolution operator
${\cal A}\cdot F(X_t) =g F'(X_t)+\frac{1}{2}F''(X_t)$. This
indeed corresponds to the stochastic equation $dX_t=d\hat B_t +g dt$.
This follows from direct computation using the It\^o derivative
$M_t^{-1}d(F(X_t)M_t)=(g F(X_t)+ \frac{1}{2}F'(X_t))dB_t+
(g F'(X_t)+\frac{1}{2}F''(X_t))dt$. 

This na\"{\i}ve example can also serve as a warning. It is well known
that the percolation interface on a domain cut in the hexagonal
lattice can be constructed as an exploration process. If the beginning
of the interface is constructed, its last step separates two hexagons
of different colors, and its end touches a third hexagon. Either this
third hexagon has already been colored or one tosses a coin to decide
the color, and then the path makes another step along an edge
separating two hexagons of different colors. The interface ends when
it exits the domain. Hence one can encode each interface by a coin
tossing game (of random duration). If the domain is the upper half
plane, the length of the game is always infinite, and there is a
simple one to one correspondance between percolation interfaces and
random walks (simple in principle, there is some subtlety hidden in
the fact that sometimes one can make one or several interface steps
without the need to toss a coin, so that the number of steps of the
interface is not simply related to the number of coin
tossings). Critical percolation corresponds to the simple symmetric
random walk with $w=1$, $v=0$. As recalled in the main text, the
scaling region for critical percolation leads to a scaling
$v \sim a^{3/4}$. On the other hand, the scaling region for the
random walk is $v\sim a \ll a^{3/4}$. This means that if one
uses a random walk with a (non critical) scaling limit, the
corresponding percolation interface is still critical, and
symmetrically that if one looks at a percolation interface in the (non
critical) scaling limit, the corresponding random walk is not
described by the scaling region.

\bigskip

\bigskip

\bigskip

\emph{Aknowledgements:}
We wish to thank Vincent Beffara, Julien Dub\'edat, Greg Lawler, Stas Smirnov and Wendelin Werner for discussions at various stages of this work. 

Our work is supported by
ANR-06-BLAN-0058-01 (D.B.), ANR-06-BLAN-0058-02 (M.B.) and ENRAGE
European Network MRTN-CT-2004-5616 (M.B. and K.K.).

\end{document}